\documentclass{mn2e}
\usepackage{color}
\usepackage{graphicx}
\usepackage{dcolumn}
\usepackage{bm}
\begin{document}

\title[Massive Black Holes in Dense Nuclear Star Clusters]
{Quantifying the Coexistence of Massive Black Holes and Dense Nuclear Star Clusters}

\author[Graham \& Spitler]
{Alister W.\ Graham$^{1}$\thanks{AGraham@astro.swin.edu.au}
and
Lee R.\ Spitler$^{1}$\\
%
$^1$ Centre for Astrophysics and Supercomputing, Swinburne University
of Technology, Hawthorn, Victoria 3122, Australia.\\
%
}

\date{Accepted 2009 May 19}

\input{psfig}

\maketitle
\label{firstpage}

\begin{abstract}

In large spheroidal stellar systems, such as elliptical galaxies, one
invariably finds a $10^6$-$10^{9} M_{\odot}$ supermassive black hole at
their centre. In contrast, within dwarf elliptical galaxies one
predominantly observes a $10^5$-$10^7 M_{\odot}$ nuclear star cluster.  To
date, few galaxies have been found with both type of nuclei coexisting and even less
have had the masses determined for both central components.  Here we
identify one dozen galaxies housing nuclear star clusters and 
supermassive black holes whose masses have been measured. 
This doubles the known number of such 
hermaphrodite nuclei --- which are expected to be fruitful sources of
gravitational radiation.  Over the host spheroid (stellar) mass range $10^8$--$10^{11}
M_{\odot}$, we find that a galaxy's nucleus-to-spheroid (baryon) 
mass ratio is not a constant value but decreases 
from a few percent to $\sim$0.3 percent such that $\log [ (M_{\rm BH} + M_{\rm
NC})/M_{\rm sph} ] = -(0.39\pm0.07)\log [ M_{\rm sph}/10^{10} M_{\odot} ]
-(2.18\pm0.07)$.  Once dry merging has commenced by $M_{\rm sph} \approx
10^{11} M_{\odot}$ and the nuclear star clusters have disappeared, this
ratio is expected to become a constant value. 
%

As a byproduct of our investigation, we have found that the projected flux
from resolved nuclear star clusters can be well approximated with S\'ersic
functions having a range of indices from $\sim$0.5 to $\sim$3, the latter
index describing the Milky Way's nuclear star cluster.
%

\end{abstract}


\section{Introduction}

Ferrarese et al.\ (2006a) and Wehner \& Harris (2006) have recently
shown that the division between either a (black hole)- or a (nuclear
cluster)-dominated galaxy core occurs around a galaxy mass of $\sim$10$^{10}
M_{\odot}$.  Wehner \& Harris (2006) wrote that ``dE,N nuclei themselves [the
nuclear star clusters] show no evidence of harbouring massive black holes''.
Indeed, in the contemporaneous investigation by Ferrarese et al.\ (2006a),
they identified only two galaxies (M32 and the Milky Way) as
potentially hosting both types of nuclear component.
While apparently rare, Filipenko \& Ho (2003) had identified at least one
galaxy with both a nuclear cluster (NC) and a massive black hole (BH) and
Graham \& Driver (2007) subsequently reported on the existence of two
additional such galaxies (NGC~3384 and NGC~7457).

To investigate this near dichotomy in the type of central massive object which
galaxies house, Seth et al.\ (2008) searched for evidence of active galactic
nuclei (AGN), and thus massive BHs, in galaxies with known NCs.  Gonzalez
Delgado et al.\ (2008) simultaneously undertook a complementary approach and
searched for the presence of NCs in galaxies with known AGN.  While they both
detected some galaxies in the mass range $10^9$ to $10^{11} M_{\odot}$ which
contain both a NC and an AGN, they were not able to acquire the
BH masses of the AGN.

To explore not only how commonplace these systems are, but importantly the
nature of the above mentioned transition, we have searched for NCs in galaxies
whose BH mass has already been determined via direct dynamical measurements.
This is important because the nucleus-to-(host spheroid) mass ratio, as a
function of spheroid mass, may provide useful constraints for potential
galactic evolutionary assembly processes.  For example, some massive BHs may
grow through the runaway collision of NC stars (e.g., Lightman \& Shapiro
1978; Kochanek et al.\ 1987; Lee 1993), or conversely the BH may evaporate the
surrounding NC (e.g., Ebisuzaki et al.\ 2001; O'Leary et al.\ 2006), or
perhaps some other mechanism dominates. 
Curiously, the continuous relations shown by Wehner \& Harris (2006) and 
Ferrarese et al.\ (2006a) involving either the BH or NC mass and the 
host galaxy mass are suggestive of, at some level, mutually common physics governing
the two types of nuclei. 
Moreover, given that the main mechanism of galaxy growth is thought to be
through the process of hierarchical merging, modelling their dual nuclei may 
be important for properly understanding the growth of supermassive black holes. 
For example, dense nuclear star clusters may, through N-body interactions, greatly
facilitate the coalesence rate of binary massive black holes.

The coexistence of NCs and massive BHs is of further interest due to
associated physical phenomenon.  The inward spiral of stars 
onto a massive BH is a likely source of UV/X-ray flaring events
(Komossa \& Bade 1999; Komossa \& Merritt 2008; Lodato et al.\ 2008; Rosswog
et al.\ 2008). 
The disruption of binary stars may result in the high-speed ejection of
hypervelocity stars (Bromley et al.\ 2006).  
Rapid inspiral 
events may also generate gravitational radiation (e.g., Quinlan 1996;
Alexander 2008; O'Leary et al.\ 2008; Merritt 2008).
As such, galactic nuclei with confirmed BHs 
and NCs may prove useful targets for experiments such as the Laser
Interferometer Space Antenna (LISA, Danzmann et al.\ 1996) 
which are hoping to discover such as yet undetected radiation.  Given that the amplitude
of gravitational waves decays linearly with distance, and X-ray flaring events
with 
distance squared, the relative proximity
of the galaxies listed here makes them particularly attractive compared
to more distant, nucleated AGN. 

In the following section we briefly describe our galaxy data set, with more
detailed information contained within the Appendix.  In Section 3 we present a
tentative new scaling law involving dense star clusters and massive black
holes, plus a more robust scaling relation which involves the mass of the host
spheroid.  Finally a discussion, including some of the implications of
this work, and a brief summary are provided in Sections~4 and 5 respectively.

\section{Data}

Graham (2008a) tabulated a sample of 50 (+26) predominantly inactive galaxies
with useful (rough) measurements of their central BH mass.  
This compilation was acquired by scouring the literature for published
values which were then  updated if new distances were available. Of these 76 galaxies,
Table~\ref{Tab1} lists those which additionally
contain a NC.  Included in this list is our own galaxy the Milky Way 
(Burbidge 1970; Rubin 1974), 
M32 (Tonry 1984), 
the active galaxies NGC~3621 and NGC~4395 (Barth et al.\ 2008; 
Filippenko \& Ho 2003), plus NGC~3384 and NGC~7457 which were previously noted
to contain both types of nuclei. 
An additional seven predominantly 
inactive galaxies (NGC 1023, 1399, 2778, 3115, 4026\footnote{Taken from
  G\"ultekin et al.\ (2009).}, 4564. 4697) which house both
type of nuclear component have been identified 
 --- although no NC mass is currently available for NGC~4564.  
In addition, Table~\ref{Tab1} includes another four galaxies with known NC
masses but only upper limits on their BH masses (as is the case for NGC~3621
mentioned above), 
three globular clusters with
possible BHs, twelve core galaxies with no NC (included for reference) and one
young star cluster (MGG-11) with a probable intermediate mass black hole.
%
The massive globular clusters included here are of interest for scenarioes in 
which they may be the relic nuclei of stripped galaxies (e.g.\ 
Freeman 1993; Bassino et al.\ 1994; Meylan et al.\ 2001; Bekki et al.\ 2003;
Walcher et al.\ 2005). 

The Appendix contains references to, or derivations of, all
quantities shown in Table~\ref{Tab1}.  
Briefly, the black hole masses have been taken from the individual
(usually discovery) papers which reported these values, and, when necessary,
adjusted to our updated distances which are also provided in the Appendix.
While galaxy masses were used for the elliptical galaxies, bulge masses have been
used for the disc galaxies.  From here on we shall generically refer to an
elliptical galaxy or the bulge of a disc galaxy as a ``spheroid''.  The
spheroid masses were primarily obtained by multiplying the observed spheroid
luminosity by an appropriate stellar mass-to-light ($M/L$) ratio.  The NC
masses were also obtained this way, albeit using a different stellar
mass-to-light ratio from the spheroid's value.  We found that the uncertainty
involved in this process is generally constrained to within a factor of two.
While comparison with dynamically-determined masses, when available, supports
this level of accuracy, detailed spectroscopy that establishes the mean ages
and metallicities of the stars (e.g.\ Walcher et al.\ 2005) is desirable
for better constraining $M/L$ ratios.  Such details, however, were not available for
most systems and therefore we usually adopted the single colour approach used by
Ferrarese et al.\ (2006a) and Seth et al.\ (2008) to determine the nuclear cluster
masses.  For five galaxies (M32, NGC~205, NGC~2778, NGC~4697 and the Milky
Way) we have modelled, in the Appendix, their observed light distribution to 
derive their NC fluxes.
%

\begin{table}
\caption{Black hole, host spheroid and nuclear cluster mass.}
\label{Tab1}
\begin{tabular}{@{}llccc@{}}
\hline
Object & Type & $M_{\rm BH} [M_{\odot}]$  &  $M_{\rm sph} [M_{\odot}]$  &  $M_{\rm NC} [M_{\odot}]$  \\
\hline
\multicolumn{5}{c}{Twelve ``core galaxies'' with $M_{\rm BH}$ but no detectable NC} \\
NGC 3379         & E    &  $1.4^{+2.7}_{-1.0}\times10^8$  &  $1.0\times10^{11}$ &      ...           \\
NGC 3608         & E    &  $1.9^{+1.0}_{-0.6}\times10^8$  &  $9.4\times10^{10}$ &      ...           \\
NGC 4261         & E    &  $5.2^{+1.0}_{-1.1}\times10^8$  &  $3.7\times10^{11}$ &      ...           \\
NGC 4291         & E    &  $3.1^{+0.8}_{-2.3}\times10^8$  &  $7.8\times10^{10}$ &      ...           \\
NGC 4374         & E    &  $4.6^{+3.5}_{-1.8}\times10^8$  &  $4.1\times10^{11}$ &      ...           \\
NGC 4473         & E    &  $1.1^{+0.4}_{-0.8}\times10^8$  &  $7.1\times10^{10}$ &      ...           \\
NGC 4486         & E    &  $3.4^{+1.0}_{-1.0}\times10^9$  &  $3.7\times10^{11}$ &      ...           \\
NGC 4649         & E    &  $2.0^{+0.4}_{-0.6}\times10^9$  &  $4.5\times10^{11}$ &      ...           \\
NGC 5077         & E    &  $7.4^{+4.7}_{-3.0}\times10^8$  &  $1.1\times10^{11}$ &      ...           \\
NGC 5813         & E    &  $7.0^{+1.1}_{-1.1}\times10^8$  &  $1.4\times10^{11}$ &      ...           \\
NGC 6251         & E    &  $5.9^{+2.0}_{-2.0}\times10^8$  &  $9.4\times10^{11}$ &      ...           \\
NGC 7052         & E    &  $3.7^{+2.6}_{-1.5}\times10^8$  &  $1.7\times10^{11}$ &      ...           \\
\multicolumn{5}{c}{Twelve galaxies with $M_{\rm BH}$ and a NC} \\ 
Milky Way        & SBbc &  $3.7^{+0.2}_{-0.2}\times10^6$  &  $1.2\times10^{10}$ &  $3.0\times10^7$   \\
M32              & cE   &  $2.5^{+0.5}_{-0.5}\times10^6$  &  $2.6\times10^{8}$  &  $2.0\times10^7$   \\
NGC 1023         & SB0  &  $4.4^{+0.5}_{-0.5}\times10^7$  &  $3.2\times10^{10}$ &  $4.4\times10^6$   \\
NGC 1399$^a$     & E    &  $4.8^{+0.7}_{-0.7}\times10^8$  &  $1.5\times10^{11}$ &  $6.4\times10^6$   \\
NGC 2778$^b$     & SB0  &  $1.4^{+0.8}_{-0.9}\times10^7$  &  $4.3\times10^9$    &  $6.7\times10^6$   \\
NGC 3115         & S0   &  $9.1^{+9.9}_{-2.8}\times10^8$  &  $7.4\times10^{10}$ &  $1.5\times10^7$   \\
NGC 3384         & SB0  &  $1.6^{+0.1}_{-0.2}\times10^7$  &  $1.4\times10^{10}$ &  $2.2\times10^7$   \\
NGC 4026         & S0   &  $1.8^{+0.6}_{-0.4}\times10^8$  &  $9.6\times10^9$    &  $5.6\times10^6$   \\
NGC 4395$^c$     & Sm   &  $3.2^{+6.8}_{-2.2}\times10^4$  &  $3.4\times10^7$    &  $1.4\times10^6$   \\ 
NGC 4564         & S0   &  $5.6^{+0.3}_{-0.8}\times10^7$  &  $7.4\times10^{9}$  &        ?           \\
NGC 4697         & E    &  $1.7^{+0.2}_{-0.1}\times10^8$  &  $1.5\times10^{11}$ &  $2.8\times10^7$   \\
NGC 7457$^d$     & S0   &  $3.5^{+1.1}_{-1.4}\times10^6$  &  $1.1\times10^{9}$  &  $9.3\times10^6$   \\
\multicolumn{5}{c}{Galaxies with a NC but only an upper limit on $M_{\rm BH}$}   \\
M33              & Scd  &  $<3\times10^3$                 &  $1.5\times10^8$    &  $2\times10^6$     \\
NGC 205          & E    &  $<2.4\times10^4$               &  $8.7\times10^8$    &  $1.4\times10^6$   \\
NGC 3621         & Sd   &  $<3.6\times10^4$               &  $1.4\times10^8$    &  $1.0\times10^7$   \\
NGC 4041$^e$     & Sbc  &  $<2.4\times10^7$               &  $6.4\times10^8$    &  $2.9\times10^7$   \\ 
VCC 1254         & dE   &  $<9\times10^6$                 &  $3.2\times10^9$    &  $1.1\times10^7$   \\
\multicolumn{5}{c}{Star clusters with less secure $M_{\rm BH}$} \\
G1$^f$           & GC   &  $1.8^{+0.5}_{-0.5}\times10^4$  &      ...            &  $8.0\times10^6$   \\
M15$^f$          & GC   &  $0.5^{+2.5}_{-0.5}\times10^3$  &      ...            &  $7.0\times10^5$   \\
MGG-11$^c$       & SC   &  $1.0^{+4.0}_{-0.8}\times10^3$  &      ...            &  $3.5\times10^5$   \\
$\omega$ Cen$^f$ & GC   &  $4.0^{+0.8}_{-1.0}\times10^4$  &      ...            &  $4.7\times10^6$   \\
\hline
\end{tabular} 

\noindent
References are available in the extended version of this Table, provided in the 
Appendix.  Uncertainties on the spheroid stellar masses and the nuclear cluster 
masses are roughly a factor of two.  The three globular clusters and one young 
star cluster (MGG-11) have no associated spheroid mass as they are not 
located at the centre of a spheroid. 
Notes: 
$^a$ NC detection weak; $^b$ NC \& BH detection weak; 
$^c$ indirect BH mass estimate; $^d$ BH detection weak; 
$^e$ disc might be dynamically decoupled; 
$^f$ maybe no BH.
\end{table}

Although roughly one dozen galaxies (including NGC~3621 and excluding NGC~4026) 
from the sample of 76 galaxies in Graham (2008a) appear to have both a NC and
a BH, it would be inappropriate to conclude that roughly 16 per cent of galaxies
contain both.  This is because sample selection effects have not been
considered.  For example, high mass galaxies tend not to have nuclear star
clusters; a sample dominated by such galaxies would be biased toward low
percentages.  In passing we note that because the central stellar density in
high mass, bright elliptical galaxies decreases as a function of increasing
galaxy luminosity (e.g., Faber et al.\ 1997), nuclear star clusters are
actually easier to detect in luminous galaxies than in intermediate luminosity
elliptical galaxies.  At the other end of the scale, the sphere-of-influence
of a $10^6 M_{\odot}$ BH within a lower-mass spheroid having a velocity
dispersion of 100 km s$^{-1}$ is only $\sim$0.01 arcseconds at the distance of
the Virgo galaxy cluster; such BHs would therefore go undetected.
Gallo et al.\ (2008) have however reported that 3-44 per cent of early-type
galaxies less massive than $10^{10} M_{\odot}$ have an X-ray active BH, while
49-87 per cent of more massive early-type galaxies do.  This may in part be a
reflection that massive BHs are less prevalent in lower mass galaxies.  In any
event, our galaxy identification in Table~\ref{Tab1} confirms that the
coexistence of NCs and BHs is not as rare as previously thought.  
Table~\ref{Tab1} effectively doubles the number of galaxies reported 
to contain a dense nuclear star cluster and having 
a direct supermassive black hole mass measurement. 

Shown in Table~\ref{Tab1} is the morphological type of each object.  Not
surprisingly, the first dozen galaxies with a BH but no signs of a NC are big
elliptical galaxies.  The next dozen objects, those with evidence for both a
BH and a NC, are predominantly disc galaxies; the exceptions are the
elliptical galaxy NGC~4697, the ``compact elliptical'' galaxy M32 (which may
be a disc galaxy undergoing transformation, e.g.\ Bekki et al.\ 2001; Graham
2002) and the elliptical galaxy NGC~1399 with only tentative evidence for a NC
(possibly a swallowed GC, Lyubenova et al.\ 2008, which is one of the propsed 
mechanisms for building NCs).
However, given that almost every galaxy with a reliable BH mass measurement
that is less than $5\times10^7 M_{\odot}$ is a disc galaxy, their prevalence is not
surprising. 
Finally, lacking kinematical information on the level of rotational
versus pressure support in the bulges of our sample, we are unable to comment
on the role that pseudobulges versus classical bulges may play.

In the following section we attempt to probe the nature of the transition
from one type of nuclei to the other.

\section{Mass ratios} 

\subsection{From star clusters to massive black holes}

Figure~\ref{Fig_BHNC} shows the ratio of the BH mass to the combined BH plus
NC mass.  It is plotted against the stellar mass of the host spheroid: either an
elliptical galaxy, the bulge of a disc galaxy, or nothing in the case of the
three globular clusters and one young star cluster (see Table~\ref{Tab1}).
For spheroids
with stellar masses below 
$\sim$$10^8 M_{\odot}$ there is a dearth of reliable BH detections, although the
majority of low-mass spheroids are known to contain NCs (e.g., 
Binggeli et al.\ 1987; Ferguson 1989; 
Carollo et al.\ 1998; Stiavelli et al.\ 2001; Balcells et al.\ 2003; Graham \&
Guzm\'an 2003; C\^ot\'e et al.\ 2006). 
From Local Group dwarf galaxies, such as NGC~205, we know that any potential
BHs which these low mass galaxies might host are less massive than their NCs.  This is
reflected by the upper limits on five of the data points in Figure~\ref{Fig_BHNC}. 
The situation is reversed for spheroid masses greater than
$\sim 10^{11} M_{\odot}$, where the BHs dominate at the expense of the NCs.
Figure~\ref{Fig_BHNC} reveals that 
in between is mutual ground where both BHs and NCs appear to coexist within
the same spheroid. 
For the first time we are able to gain some preliminary insight into the
nature of this transition as a function of mass, although we recognise that
more data is needed in Figure~\ref{Fig_BHNC} before any possible relation can
be defined with certainty.

The demise of NCs at a host spheroid mass of $\sim 10^{11} M_{\odot}$
(Figure~\ref{Fig_BHNC}, see also Ferrarese et al.\ 2006a and Wehner \& Harris
2006) is interesting.  The onset of partially depleted galaxy cores occurs at
an absolute $B$-band magnitude of $-20.5\pm1$ mag (e.g., Faber et al.\ 1997;
Graham \& Guzm\'an 2003), which is also where the dynamical properties vary 
(e.g., Davies et al.\ 1983; Dressler \& Sandage 1983; Matkovi\'c \& Guzm\'an
2005).  For an old stellar population, this stellar flux corresponds to a
stellar mass of $6^{+9}_{-4}\times10^{10} M_{\odot}$ --- which has recently
been noted by many studies as marking the transition of several galaxy
properties (e.g.\ Rogers et al.\ 2008) and may also coincide with the turnover of the
galaxy mass function (Li \& White 2009).  As noted by Ferrarese et al.\
(2006a), it may therefore be that coalescing 
BHs in dry merger events (Begelman, Blandford, \& Rees 1980; Merritt, Mikkola
\& Szell 2007; Berentzen et al.\ 2009) preferentially destroy their shroud of
NC stars prior to the creation of the galactic loss cones observed in
spheroids brighter than $-20.5\pm1$ $B$-mag.  Alternatively, perhaps the life
span of a NC is simply short once the mass of the BH dominates, hence the
scarcity of NCs around BHs with $M_{\rm bh} > \sim5 \times 10^7 M_{\odot}$.

Figure~\ref{Fig_BHBH} shows the same mass ratio as seen in Figure 1, but
plotted against the BH mass.  Plotting it like this reveals, without recourse
to the host spheroid, the nature of the coexistence of black holes and dense
star clusters.  The line shown in Figure~\ref{Fig_BHBH} has simply been 
marked by eye to roughly capture the behaviour of the points and is such that
\begin{equation}
\log \left[ \frac{M_{\rm BH}}{M_{\rm BH}+M_{\rm NC}} \right] = \frac{2}{3}
\log \left[ \frac{M_{\rm BH}}{5\times 10^7 M_{\odot}} \right]
\label{Eq_Rat} 
\end{equation}
for $M_{\rm BH} < 5\times 10^7 M_{\odot}$ and equals zero for larger BH
masses and when $M_{\rm NC}=0$.  Given the somewhat sparse nature of the data,
a more sophisticated regression analysis for this new (black hole)-(nuclear
cluster) mass ratio relation is not performed here.
%

\subsection{Nuclei-to-spheroid mass ratios}

It is generally accepted that massive BHs are associated with the
host spheroid rather than the host galaxy (e.g., Kormendy \& Gebhardt 2001). 
Given this, we have 
displayed in Figure~\ref{Fig_Ratio} the combined mass of the BH and the NC,
divided by the stellar mass of the host spheroid.  
For high spheroid masses, where supermassive BHs dominate 
the core region, one can see that this ratio scatters between values from 
$10^{-3}$ to $10^{-2}$.  One can also see that this mass ratio is greater in
the lower mass spheroids whose cores are dominated by a NC\footnote{If a 
population of yet-to-be-detected, low-mass spheroids with $M_{\rm BH} >
M_{\rm NC}$ exists, they would act to increase the distribution of points at the low-mass end
of Figure~\ref{Fig_Ratio} to higher values and thereby steepen the relation
further.}. 
From an orthogonal regression analysis, using the code BCES (Akritas \& Bershady 1996), 
and assuming a factor of two uncertainty on 
each data point in both directions, one obtains the relation\footnote{Excluding 
NGC~205 from Figure~\ref{Fig_Ratio} gives a consistent slope and intercept of 
$-0.41\pm0.06$ and $-2.13\pm0.07$, respectively.}
\begin{equation}
\log \left[ \frac{M_{\rm BH} + M_{\rm NC}}{M_{\rm sph}} \right] =
- (0.39\pm0.07) \log \left[ \frac{M_{\rm sph}}{10^{10} M_{\odot}} \right] -(2.18\pm0.07).
\label{Eq_Nat}
\end{equation} 
Repeating the analysis while assigning a factor of 5 uncertainty to the
ordinate (and a factor of 2 in the abscissa) does not 
change this result by more than the quoted 1$\sigma$ uncertainties. 
Setting $M_{\rm BH}=0$ for the systems which only have upper limits on their
BH masses also does not significantly alter these results.
The Pearson and Spearman correlation coefficients are -0.73 and -0.65, and the 
probability of such a strong correlation occurring by chance is less than 0.02
per cent.  
The vertical scatter (i.e.\ in the $\log M_{\rm nuclear}$ direction) is 0.41 dex,
and 0.36 dex without NGC~205.

While equation~\ref{Eq_Rat} expressed the relevant dominance of the BH
compared to the NC as a function of BH mass, equation~\ref{Eq_Nat} reveals
their combined importance (in terms of mass) relative to the host
spheroid's stellar mass. 
It also effectively provides a new means to predict the central mass in
systems where one is unable to directly measure this quantity. 

Once dry merging commences at $M_B \approx -20.5$ mag (e.g.\ Graham \&
Guzm\'an 2003, and references therein), or roughly $M_K \approx -24$ mag (or
$5\times10^{10}$ to $10^{11} M_{\odot}$), the $M_{\rm BH}/M_{\rm sph}$ mass ratio
should remain constant (or decrease if BHs can be ejected, e.g., Merritt et
al.\ 2004; Gualandris \& Merritt 2008).  The one-to-one $M_{\rm
BH}$-$L_K$ relation given by Graham (2007, which is dominated by systems with
$M_{\rm BH} > 5\times10^7 M_{\odot}$), coupled with a near constant stellar
$M/L_K$ ratio for massive elliptical galaxies, supports the scenario in which
the $M_{\rm BH}$/$M_{\rm sph}$ baryon\footnote{This terminology assumes that
the black holes have been built by baryons (e.g.\ Shankar et a.\ 2004).}  mass
ratio is a roughly constant value.
Using the $K$-band stellar mass-to-light ratio $\log (M/L_K) = 0.1-0.1(B-K)$, for 
$(B-K) > 2.3$ 
(Forbes et al.\ 2008, their Figure~10), and the colour-magnitude relation
$(B-K) = 0.082 - 0.155M_K$, for $M_K < -18$ mag
(Forbes et al.\ 2008, their Equation~1), one has the expression 
$\log (M/L_K) = 0.01(9.18 + 1.55M_K)$. 
Applying this to the $K$-band expression $M_{\rm BH} \propto L_{\rm sph,
stellar}^{1.00\pm0.05}$ from Graham (2007, his section~5.2) gives
$M_{\rm BH} \propto M_{\rm sph, stellar}^{1.04\pm0.05}$. 
%
For comparison, Marcomi \& Hunt (2003) report $M_{\rm BH} \propto M_{\rm sph,
  virial}^{0.96\pm0.07}$ for their ``Group 1'' galaxies,
while H\"aring \& Rix (2004) report $M_{\rm BH} \propto M_{\rm sph,
  dyn}^{1.12\pm0.06}$ for a slightly larger galaxy sample. 
It should however be noted that these latter two studies have,
at some level, also accounted for the contribution from dark matter 
in the spheroid mass, an isue we discuss in the following Section.

\section{Discussion}

Only a few years ago it was generally believed that massive black holes and
nuclear star clusters did not (frequently) coexist at the centres of galaxies.
Here, as in Seth et al.\ (2008) and Delgado et al.\ (2008), we present
evidence suggesting the contrary for galactic spheroids with stellar masses
ranging from $\sim10^8$--$10^{11} M_{\odot}$.  Furthermore, we take an
important step forward by reporting on systems for which we have been able to
acquire the (black hole and stellar) masses of the nuclear components and the
(stellar) mass of the host spheroid.  This has enabled us to present mass
relations defining this exciting coexistence.

From Equation~\ref{Eq_Nat}, when $M_{\rm sph} = 10^8 M_{\odot}$ one has a
nucleus-to-spheroid mass ratio of 0.04, and when $M_{\rm sph} = 10^{11}
M_{\odot}$ one has a ratio of 0.0027.  Perhaps not surprisingly, the latter
value is in excellent agreement with the mean $M_{\rm BH}/M_{\rm sph}$ ratio
from studies of galaxies at the high-mass end with $M_{\rm BH} \sim 10^{8\pm1}
M_{\odot}$, and which excluded any NC mass component (e.g.\ Merritt \&
Ferrarese 2001; H\"aring \& Rix 2004).  At
the low-mass end, from an analysis of nuclear star clusters in dwarf
elliptical galaxies and the bulges of early-type disc galaxies, the $M_{\rm
NC}/M_{\rm sph}$ stellar mass ratio has been observed to be both higher and to
increase (decrease) as one samples lower (higher) mass spheroids.  Balcells et
al.\ (2003) find a value of $\sim$2 per cent when $M_{\rm sph} = 10^8
M_{\odot}$ and from a sample of dwarf elliptical galaxies a value of 1 per
cent when $M_{\rm sph} = 10^8 M_{\odot}$ is readily derived from Graham \&
Guzm\'an (2003, their Eq.3 assuming an {\it F606W} filter mass-to-light ratio of 3).

As noted above, in low-mass spheroids it has been known for some years how the
stellar flux ratio of the nucleus and host spheroid vary (see also Lotz et
al.\ 2004, their Figure 7).  Grant et al.\ (2005), for example, report that their $B$-band
data for dwarf elliptical galaxies yields $L_{\rm NC} \propto L_{\rm
sph}^{0.68}$, which implies a nine-fold variation in the nuclear-to-spheroid
flux ratio over a host spheroid flux range of 1000.  (For comparison, given
that $M_{\rm BH}/M_{\rm NC}\approx 0$ at the low mass end of
equation~\ref{Eq_Nat}, one has the (stellar mass) relation $M_{\rm NC} \propto M_{\rm
sph}^{0.61\pm0.07}$ when the NCs dominate.  From Shen et al.'s (2008) 
analysis of 900 broad line AGN, they report that $M_{\rm BH} \propto 
L_{\rm galaxy}^{0.73\pm0.05}$.)
In Figure~\ref{Fig_Ratio} we have revealed, over a host spheroid stellar mass
range of $10^4$, how the combined central object mass (black hole plus
nuclear star cluster) divided by the stellar mass of the host spheroid varies
with the latter quantity.  This ratio increases by more than an order of
magnitude from $\sim$0.1 per cent in giant elliptical galaxies dominated by
massive black holes, to 5--9 percent in dwarf galaxies and the bulges of
late-type disc galaxies whose inner regions are dominated by a nuclear star
cluster (see also Balcells et al.\ 2007). 

At first glance, this result may appear to contradict recent claims of a
constant (central massive object)-to-(host galaxy) mass ratio, where the
central massive object in such works was either a nuclear star cluster or a
massive black hole.  In the case of Wehner \& Harris (2006), they effectively
took the above flux relation for dwarf elliptical 
galaxies from Grant et al.\ and used the expression
($M_{\rm total}/L)_{\rm sph} \propto L_{\rm sph}^{-0.3}$ to obtain $L_{\rm NC}
\propto M_{\rm sph,\, total}$.  If the nuclear clusters have similar stellar
$M/L$ ratios, this leads to the result that the nuclear cluster mass is
linearly proportional to the {\it total} (dark matter plus stellar) mass of the host
spheroid; i.e.\ that this mass ratio is constant with varying spheroid mass.
The mass ratios presented by Ferrarese et al.\ (2006a), while also accounting 
for dark matter at some level, are slightly different due to their
inclusion/treatment of disc galaxies from a sample of early-type,
Virgo cluster galaxies.  Although their application of the virial theorem ($M
\propto R_{\rm e}\sigma^2$) using the velocity dispersion $\sigma$ of the
(pressure supported) bulge component together with the effective half-light radii of the whole galaxy 
--- which are effected by the size of the (rotationally supported) disc component --- is a 
questionable meausure of a lenticular 
galaxy's mass, it is clear that these virial products are larger than the
spheroid masses that would be obtained from the use of $R_{\rm e,\, sph}$
in such a formula.\footnote{A more
  subtle issue is that the nuclear cluster fluxes (and thus masses) may have
  been underestimated due to the steeper inner S\'ersic profiles obtained from
  single higher S\'ersic index fits to each disc galaxy rather than from a
  S\'ersic bulge $+$ exponential disc fit.}  The average nucleus-to-spheroid
mass ratio would therefore be larger than the reported value of 0.2 
per cent for the nucleus-to-galaxy total mass ratio.
%
%
  
While an investigation of whether the use of $R_{\rm e}\sigma^2$ is an
appropriate tracer of total mass is beyond the intended scope of this paper,
it does seem apt to remind readers that this question has a long history,
often discussed in association with the Fundamental Plane (Djorgovski \& Davis
1987; Faber et al.\ 1987; Djorgovksi, de Carvalho \& Han 1988).  For example,
even within elliptical galaxies, luminosity-dependent dynamical non-homology
may severely bias the applicability of aperture velocity dispersion
measurements when deriving such quasi-''virial masses'' from $R_{\rm
e}\sigma^2$, and thus also bias any $M_{\rm total}/L$ trends with luminosity
(e.g.\ Hjorth \& Madsen 1995; Ciotti et al.\ 1996; Busarello et al.\ 1997;
Graham \& Colless 1997; Prugniel \& Simein 1997).
%
%
%
Aside from concerns about measuring total spheroid masses, baryonic fuelling
and feedback, albeit within a dark matter halo, are commonly thought to be
responsible for establishing the (bulk of the) nuclei mass and setting the
observed nuclear-to-(host spheroid) mass ratios (e.g.\ Silk \& Rees 1998;
Kauffmann \& Haehnelt 2000; Benson et al.\ 2003; Croton et al.\ 2006; Booth \&
Schaye 2009).  It therefore seems reasonable to construct a baryonic rather
than (only) total mass-ratio relation.  Moreover, we have been able to do so
for the first time when including the mass of both nuclear components from the
same galaxy.

While inward gas flow may result in galactic-centric star
formation or fuelling much of the growth of massive BHs (e.g.\ Shankar et al.\
2004), it has also been suggested that NCs may grow through the accretion of
globular clusters, and/or super star clusters in spiral galaxies, via 
dynamical friction (Tremaine et al.\ 1975; Quinlan \& Shapiro
1990).  Dark matter may thus also have a role to play, thereby motivating the
pursuit of reliable total masses.
It has additionally been suggested that some BHs may be built through the
runaway collision of the NC stars (Kochanek et al.\ 1987) or that,
alternatively, a massive BH may effectively evaporate the surrounding NC
(Ebisuzaki et al.\ 2001; O'Leary et al.\ 2006).
%
It has also been proffered that NCs and massive BHs may have developed from
the same initial formation process (Wehner \& Harris 2006), such that a slower
gas infall rate in smaller spheroids allows time for star formation and thus
produces nuclear star clusters rather than massive black holes.  Our larger
(stellar) mass ratios in smaller mass spheroids may have implications for the
required 
efficiency of feedback mechanisms in which supernova and stellar winds from NC
stars regulate the nuclear-to-spheroid mass ratio (McLaughlin et al.\
2006b).  Moreover, it is hoped that the nuclear-to-spheroid (baryonic) mass
ratios provided here may provide useful constraints for any potential
evolutionary scenarios.

In future work we intend to present a new diagram showing $(M_{\rm BH} +
M_{\rm NC})$ versus velocity dispersion, $\sigma$.  This will be achieved via
a careful analysis of high-resolution {\it HST} images for as many of the 50
(+26) galaxies as possible.  While barred galaxies can deviate from the
$M$-$\sigma$ relation defined by non-barred galaxies (e.g.\ Graham 2008b;
Graham \& Li 2009), and one expects them to similarly deviate in the new
$(M+M)$-$\sigma$ diagram due to their elevated values of $sigma$, it may prove
insightful to investigate this further.  If barred galaxies have
preferentially larger $M_{\rm NC}/M_{\rm BH}$ mass ratios than non-barred
galaxies of the same velocity dispersion, then one may find less scatter in
the new diagram and further potential clues to their evolution.


Within galaxy clusters, dwarf galaxies are the most common type of galaxy
(e.g.\ Binggeli et al.\ 1985) and many of these are nucleated.  Within the
field environment, the most common type of galaxies are spiral galaxies (e.g.\
Allen et al.\ 2006; Baldry et al.\ 2006) and many of these are also known to
be nucleated (e.g.\ Carollo et al.\ 1998; B\"oker et al.\ 2002; Balcells et
al.\ 2003).  Due to the difficulties associated with the detection of low mass
black holes ($<10^6 M_{\odot}$) in external galaxies, we speculate that
androgynous nuclei might be far more common than currently recognised.
Furthermore, given that nuclear star clusters are among the highest stellar
density objects in the Universe, such a commonplace coexistence of nuclear
star clusters and massive black holes may open up the prospect for numerous
detections of low frequency gravitational radiation (e.g.\ Ju et al.\ 2000,
and references therein) with the Laser Interferometer Space Antenna (LISA,
Danzmann et al.\ 1996) from rapidly inspiralling stars, white dwarfs, neutron
stars and stellar mass black holes about these massive black holes.
Due to the substantially higher density of stars in NCs, compared to the
underlying host galaxy (see Figure~\ref{Fig_Milky}), past estimates of
LISA-detectable gravitational radiation events (e.g.\ Sigurdsson 1997; 
Freitag 2001; 
Gair et
al.\ 2004; Hopman \& Alexander 2005, 2006a,b) may need to be revised upwards.
The sense of the correction is of course welcome given the significance a
direct detection could have by not only supporting Einstein's concept of space
and time but opening an entire new window through which to view, or rather
listen to, our Universe.

\section{Summary}


We have identified roughly a dozen galaxies with a direct BH mass measurement
{\it and} a nuclear star cluster, doubling the previous sample size for which
these measurements are available.  We speculate that the existence of such
hermaphrodite nuclei may be a rather common event for spheroids with
stellar-masses ranging from $10^8$ to $10^{11} M_{\odot}$ (see also Gonzalez
Delgado et al.\ 2008 and Seth et al.\ 2008).

\begin{figure*}
\includegraphics[angle=270,scale=0.93]{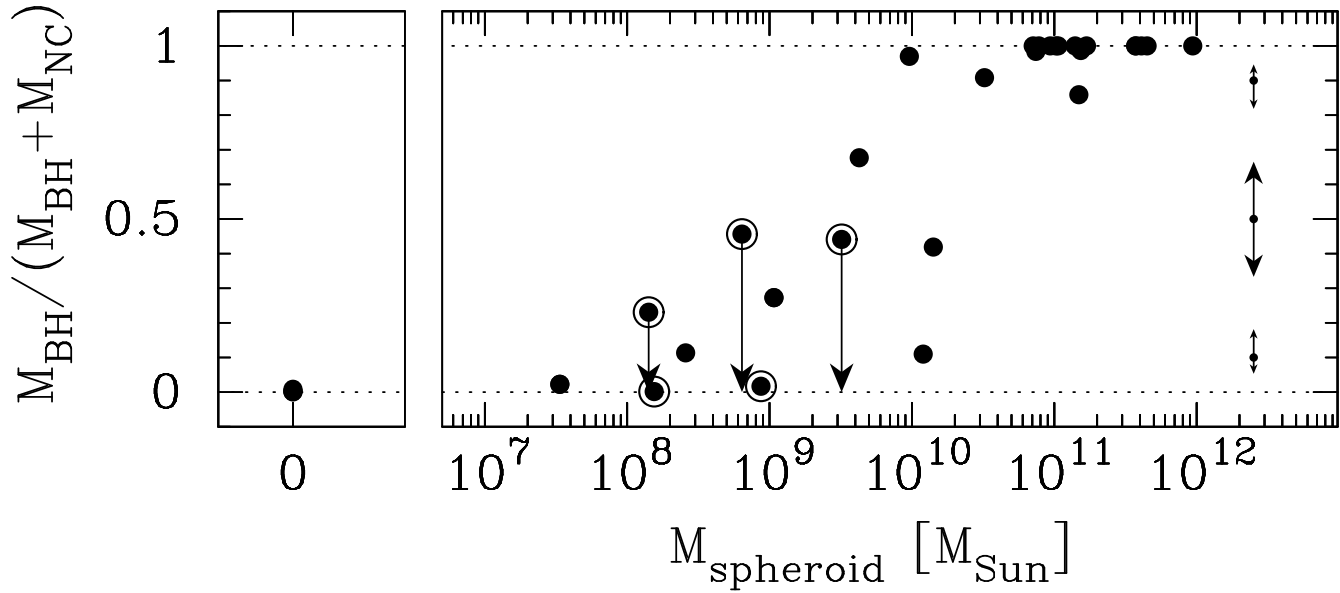}
\caption{
The increasing dominance of the central black hole over the nuclear cluster of
stars, traced by the mass ratio $M_{\rm BH} / (M_{\rm BH}+M_{\rm NC})$, 
is shown to depend on the host spheroid mass $M_{\rm sph}$ --- which is zero for GCs.
The highest mass spheroids do not contain a NC, but are included for 
illustrative purposes (see Table~\ref{Tab1}). 
The five circled points have only an upper limit to their BH mass. 
The arrows on the right hand side of the figure denote the up and down
movement if either of the two nuclear masses are in error by a factor of $\pm2$.
}
\label{Fig_BHNC}
\end{figure*}

We have shown that the mass of the nuclear component(s) increases from $\sim$0.1
per cent of the host spheroid's stellar mass in large elliptical galaxies 
whose cores are dominated by a massive black hole, 
to several per cent in low stellar mass ($\sim 10^8 M_{\odot}$) spheroids 
whose cores are dominated by a nuclear star cluster (see
Figures~\ref{Fig_BHNC} and \ref{Fig_Ratio}, and also Balcells et al.\ 2007). 

We have derived a linear relation between the nuclear mass (BH and NC
combined) and the stellar mass of the host spheroid.  Given in 
equation~\ref{Eq_Nat}, 
the relation can be expressed as $(M_{\rm BH} + M_{\rm NC}) \propto M_{*,
  sph}^{0.61\pm0.07}$ (see Figure~\ref{Fig_Ratio}). 
We hereby suggest that this baryonic, nuclear-to-spheroid mass ratio relation is 
applicable to spheroids with eith either nuclear clusters or
dual nuclei type, 
noting that the exponent may equal a value of 1 once dry merging commences 
(assuming no loss of nuclear components). 

\begin{figure}
\includegraphics[angle=270,scale=0.43]{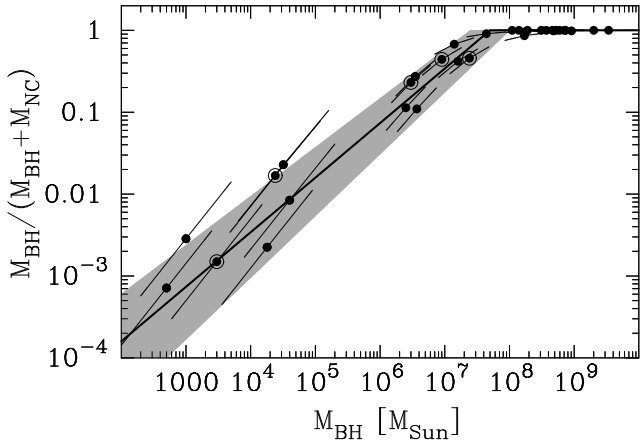}
\caption{
The mass ratio from Figure~\ref{Fig_BHNC} is shown here 
versus the mass of the black hole.  Symbols are as in Figure 1.  
An expression for the solid line is given in Equation~\ref{Eq_Rat}. 
Galaxies with only an upper BH mass limit have been circled. 
An error in $M_{\rm BH}$ will move the data points nearly parallel to this relation,
insuring that it is preserved in the presence of $M_{\rm BH}$ measurement errors.
The lines emanating from each data point show how much each point would
move if the black hole mass changed by a factor of $\pm2$ (upper points) or
$\pm5$ (lower seven points). 
Reflecting the increasing uncertainty on the smaller BH mass measurements,  
the shaded area has a horizontal width of log(2.0) at the top and
log(5.0) at the mass ratio $5\times 10^{-3}$. 
}
\label{Fig_BHBH}
\end{figure}

\begin{figure}
\includegraphics[angle=270,scale=0.44]{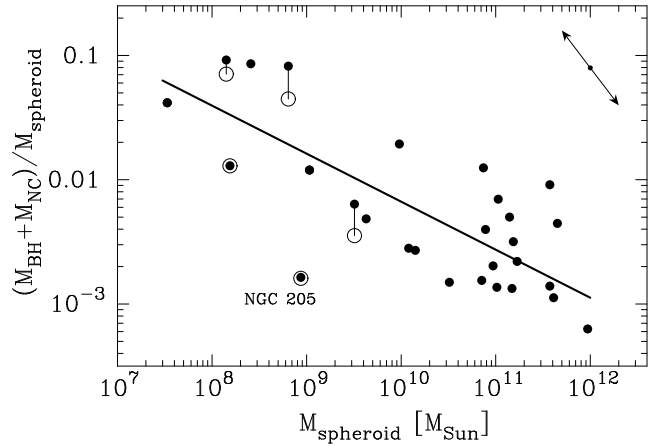}
\caption{
The importance, in terms of mass, of the nuclear components relative to the
host spheroid, as traced by the mass ratio $(M_{\rm BH}+M_{\rm NC})/M_{\rm
  spheroid}$, is shown as a function of $M_{\rm spheroid}$. 
Systems with only an upper BH mass limit have a connected circle which shows their 
location if $M_{\rm BH} = 0$. 
The fitted line is given by equation~\ref{Eq_Nat}. 
The bi-directional arrow in the top right reveals how the points would 
move if the value of $M_{\rm spheroid}$ changes by a factor of $\pm$2. 
No arrow is provided for changes in the nuclear component masses because the 
result/movement is obvious, with a simple shift in the ordinate.
}
\label{Fig_Ratio}
\end{figure}

We have also identified a new (black hole)-(nuclear cluster) mass ratio
relation pertaining to the coexistence of these entities.  We provide a
preliminary quatification of this relation in equation~\ref{Eq_Rat}, which is
such that $(M_{\rm BH} + M_{\rm NC}) = (5\times 10^7 M_{\odot})^{2/3}\, M_{\rm
BH}^{1/3}$ for $M_{\rm BH} < 5\times 10^7 M_{\odot}$ (see
Figure~\ref{Fig_BHBH}).

\section{acknowledgment}

We thank Kenji Bekki for motivating us in 2007 to undertake this project. 
A.G.\ thanks the organisers of the February 2008 ``Nuclear Star clusters
across the Hubble Sequence'' workshop held at the Max-Planck-Institut f\"ur
Astronomie in Heidelberg, Germany, and the organisers of the July 2008 Lorentz
Centre Workshop ``Central Mass Concentrations in Galaxies'' held in Leiden,
The Netherlands, where preliminary versions of Figure~\ref{Fig_BHNC} were 
presented.

\section{APPENDIX: The Data}

Here we provide sufficient information for the reader to reconstruct our data
set, shown in Tables~\ref{Tab1} and \ref{TabS1}, and include the necessary references to
relevant sources of literature.  When no clear (refereed) literature data was
available for a nuclear cluster's magnitude, we have modelled the galaxy's
light profile ourselves to determine the stellar flux.  Select notes on
individual objects are provided below, along with a brief mention of some
objects that were excluded.

Columns 1 and 2 of Table~\ref{TabS1} provide the object name and its morphological
type.  The black hole masses (column~4 of Table~\ref{TabS1}) have been adjusted
according to updated information on their distances, which are provided in
column~3.

Galaxy magnitudes were used for the elliptical galaxies, while bulge magnitudes
were used for the lenticular (and the few spiral) galaxies. 
These are listed in column~4 of Table~\ref{TabS1}. 
These ``spheroid'' magnitudes were then converted into solar masses.  
This first required that we calibrate them in terms of solar luminosity, 
and we used the following absolute solar magnitudes (Cox 2000): 
$M_B = 5.47$; 
$M_V = 4.82$; 
$M_K = 3.33$; 
supplemented with  
$M_H = 3.32$ (Bessell et al.\ 1998); 
$M_{F814W} = 4.14$ and 
$M_{F850LP} = 3.99$\footnote{http://www.ucolick.org/$\sim$cnaw/sun.html}.
These spheroid luminosities were then converted into masses using an appropriate stellar
mass-to-light ($M/L$) ratio: for elliptical galaxies (and lenticular bulges)
we assumed an age of 13 Gyr and [Fe/H] = 0.5 dex (and 7 Gyr and [Fe/H] = 0.3
dex), which gave a $K$-band stellar mass-to-light ratio of $\log [M/L_K] =
0.0$ dex (and $\log [M/L_K] = -0.22$ dex).  Given that the $K$-band stellar
mass-to-light ratio varies little with metallicities ranging from [Fe/H] = 0.0
to 0.6 dex, and with ages ranging from 5 to 13 Gyr, the above selection of age
and metallicity has little effects on the results.  For the bulges in spiral
galaxies, and when no reliable $K$-band magnitudes were available, the
derivation of the stellar $M/L$ ratio is given below.  The adopted ratios for
each passband are shown in column~5 of Table~\ref{TabS1}, while the
spheroid masses are given in column~6.   Given the various sources of
uncertainty in each step, these values are likely to be accurate to within a
factor of two.


Because the NCs may have a considerable range of ages (and metallicities),
individual stellar $M/L$ ratios were determined for each NC.  To estimate
these we have used the available NC colours (see below) together with the
models by Bruzual \& Charlot (2003) and a Chabrier (2003) initial mass
function (see Figure~\ref{Fig_Lee}).  The approach adopted is discussed below
for each NC, and the results are shown in columns~7 to 9 of Table~\ref{TabS1}. 
In some instances we bypassed this process and used literature
established masses.  
%

\subsection{Notes on individual galaxies}

{\bf Milky Way:}
Due to its proximity, the centre of our own galaxy, the Milky Way, has long
been suspected of harbouring both a nuclear star cluster and a supermassive
black hole (Rubin 1974). 
References to the masses of these entities (not the mass
within the inner parsec: Sch\"odel et al.\ 2009b) 
are provided in Table~\ref{TabS1}, 
along with an estimate of the mass of the Milky Way's bulge. 

The nuclear star cluster has most recently been examined by Sch\"odel et al.\
(2009a) and Oh et al.\ (2009). 
In Figure~\ref{Fig_Milky} we have performed a separation of the 
Milky Way's NC light from the background bulge (and at some level bar)
light using a standard decomposition technique.  For perhaps the 
first time, we report that the nuclear cluster 
of the Milky Way is reasonably well represented by a S\'ersic function 
with index $n=3$; it has an effective half-light radius $R_{\rm e}$ equal 
to 80 arcseconds (3.2 pc).  It is perhaps worth emphasising that this is
far from a Gaussian distribution which would have a S\'ersic index of 0.5.
Determination of whether alternative models, such as the King (1962) model, 
provide a better fit will be left for a separate study. 

Due to the uncalibrated nature of the published light profile, we are not 
able to report a magnitude for the nuclear cluster. However, even though 
no extinction corrections were applied, 
having accounted for the background bulge flux in Figure~\ref{Fig_Milky}, 
we can provide rough estimates 
to the negative logarithmic slope of the nuclear cluster's projected density profile 
$\gamma(R) \equiv {-{\rm d}[\log I(R)]/{\rm d}\log R} 
= (b_n/n)(R/R_{\rm e})^{1/n}$, with $b_n \approx 1.9992n - 0.3271$ (Capaccioli
1989; Graham \& Driver 2005). 
At $R=R_{\rm e}/4 =20^{\prime \prime}$, $\gamma = 1.19$, and at 
$R=R_{\rm e}=80^{\prime \prime}$, $\gamma = 1.89$.  The 
associated negative logarithmic slope of the nuclear cluster's internal 
3D (i.e.\ non-projected) density profile is 2.0 and 2.7, 
respectively (Graham et al.\ 2006, their Eq.23). 
This is steeper than previous estimates of $\sim$1.4 to 1.8 
(Becklin \& Neugebauer 1968; Catchpole et al.\ 1990; 
Sch\"odel et al.\ 2009a, 2009b and references
therein) which have been biased by bulge stars.

{\bf M32} 
Ferrarese et al.\ (2006a) suggest that M32 may contain both a nuclear star
cluster (Smith 1935; Burbidge 1970; Worthey 2004) 
and a BH (Verolme et al.\ 2002, $M_{\rm BH}=2.5^{+0.5}_{-0.5}\times10^6 M_{\odot}$). 
In Figure~\ref{Fig_M32} we present an $I$-band light profile which we have 
fitted with an inner nuclear component, plus a S\'ersic bulge and an outer 
exponential envelope/disc (see Graham 2002).  
The inner component and the main spheroid component have an apparent $F814W$ 
magnitude of 10.0 and 7.5 mag, respectively. 
The S\'ersic index $n$ of the 
inner component is 2.3 (the effective half-light radius is 1.65 arcseconds,
equal to $\sim$6 pc).
Together with the Milky Way's nuclear star cluster, this paper 
presents the first clear evidence/statement that 
nuclear excesses do not all have a Gaussian-like structure (i.e.\ $n=0.5$); 

Although we initially excluded M32 due to the somewhat unknown nature of its
central excess, spectroscopy has revealed
that the inner region of M32 does possess a different mean chemistry and age (e.g.\
Worthey 2004; Rose et al.\ 2005) to the main spheroid, although no obvious 
transitional radius is apparent.  
This difference has most recently 
been quantified by Coelho et al.\ (2009), using an inner 1.5 arcsecond slit to 
sample the nuclear region. 
Using the spread of ages and metallicties from their Table~3
gives an $I$-band, stellar $M/L$ ratio of $0.75\pm0.17$ and 
$0.92\pm0.18$ for the nucleus and main spheroid of M32. 
%

Exclusion of M32 from the analysis has no significant effect on any of the
results.

\begin{figure}
\includegraphics[angle=270,scale=0.4]{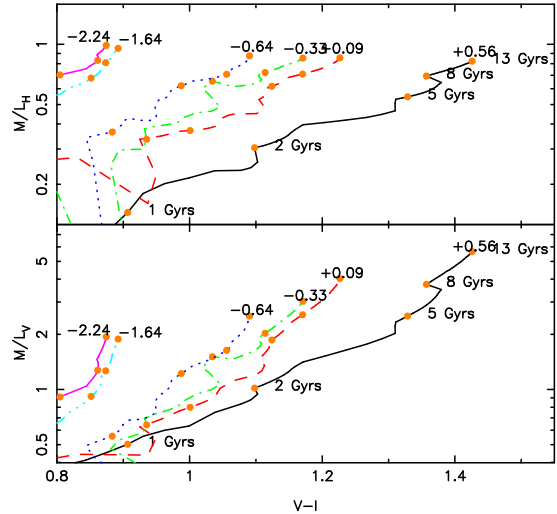}
\caption{
  $V$- and $H$-band mass-to-light ratios for a range of stellar metallicities
  ([Fe/H] = $-2.24$ to $+0.56$ dex) and ages ($<$13 Gyrs) as a function of
  $V-I$ colour.  Based on the Bruzual \& Charlot (2003) stellar population models and using a
  Chabrier (2003) initial mass function.  
}
\label{Fig_Lee}
\end{figure}

\begin{figure}
\includegraphics[angle=270,scale=0.34]{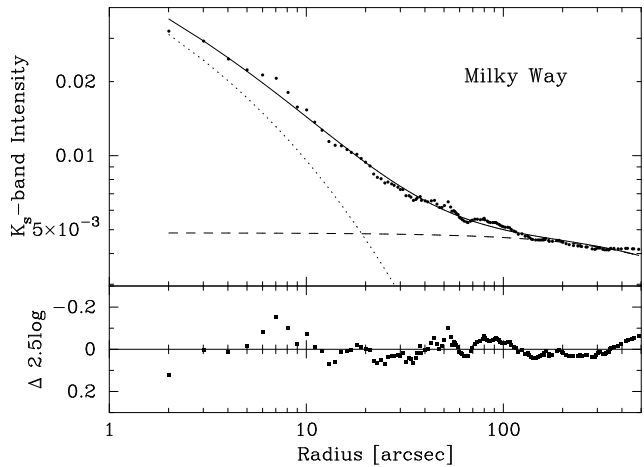}
\label{Fig_Milky} 
\caption{
Uncalibrated, 2MASS, $K_s$-band intensity profile from the centre of the 
Milky Way, taken from Sch\"odel et al.\ (2009a, their Figure 2). 
Here we have modelled the nuclear star cluster with a S\'ersic function
(dotted curve) 
and the underlying host bulge --- which has an effective half-light radius of
$\sim$4.5 degrees (e.g.\ Graham \& Driver 2007) and is therefore basically a
horizontal line --- with an exponential function (Kent et al.\ 1991).
The nuclear cluster's S\'ersic function has an index of $\sim$3.0. 
}
\end{figure}

\begin{figure}
\includegraphics[angle=270,scale=0.34]{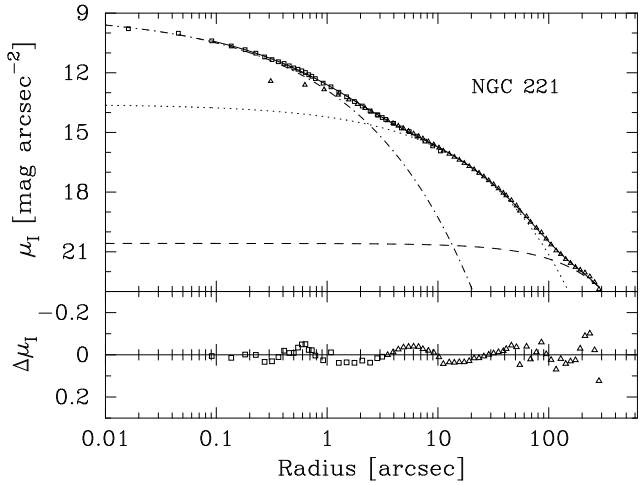}
\label{Fig_M32} 
\caption{
NGC 221 (M32). 
The inner points are the deconvolved {\it HST/WFPC2 F814W} 
(roughly Johnson $I$-band), major-axis, surface brightness
profile from Lauer et al.\ (1998), while the outer points are the $R$-band
major-axis data from Kent (1987) after a constant shift of $R-I$=0.91 
mag arcsec$^{-2}$ has been applied.
The modelled data (see the lower residual panel) has been fitted
with two S\'ersic components (a NC plus the main spheroid) 
plus an outer exponential function (see Graham 2002, and Worthey 2004). 
}
\end{figure}

{\bf NGC 1023:} 
Bower et al.\ (2001) show that the average $V-I$ colour within 1.0 (0.1)
arcsecond is 1.4 (1.25).  From the innermost colour our adopted stellar
population models tell us that the metallicity is super-solar and that the
population must be of an intermediate to old age, in agreement with the 7 Gyr
old age from Sil'chenko (1999).  From Figure~\ref{Fig_Lee}
the associated $V$-band, stellar $M/L$ ratio is in the range
$0.2 < \log [M/L_V] < 0.6$ and we adopt $M/L_V = 2.5$ for the 
NC magnitude from Bower et al.\ (2001, their Eq.2).

Faber et al.\ (1997) report $B-V=0.93$ for this galaxy, a typical colour for
an old population. 
Assuming a 13 Gyr old stellar population, we have applied a
$K$-band stellar $M/L$ ratio of 1.0 to the $K$-band bulge magnitude tabulated
in Marconi \ Hunt (2003).

%


{\bf NGC 1399:} 
While Houghton et al.\ (2006) had suggested that a nuclear disc may be present,
Gebhardt et al.\ (2007) refute the existence of any second component at 
the centre of NGC~1399.  However Lyubenova et al.\ (2008) have since suggested that 
a nuclear star cluster (perhaps a partially digested, bright
globular cluster) may reside within the inner $\sim$0.3 arcseconds (30 pc) and
we tentatively accept its existence.  If, however, NGC~1399 does not possess
a NC, it will have no impact on our results.   This is because the NC/BH 
mass ratio is so small that the data point in our Figures~\ref{Fig_BHNC} 
to \ref{Fig_Ratio} will barely move if we set $M_{\rm NC}=0$. 

For a 13 Gyr old population, $M/L_B \sim 2.5$, which was used for this galaxy. 
It is interesting to note that NGC~1399 is a ``core galaxy''.  No other 
such galaxies, with their partially depleted stellar cores, appear to 
house nuclear star clusters.
The apparent excess nuclear flux in some ``core galaxies'' (for example, the
Sy2 LINER NGC~4552) is almost invariably due to 
non-thermal radiation from an active galactic nucleus; in some instances it may
also be due to residuals from the deconvolution process used on early HST images.

{\bf NGC 2778:} 
From kinematical measurements, Rix et al.\ (1999) revealed that this object is 
a lenticular galaxy, in agreement with the designation by Kent (1985). 
Marconi \& Hunt (2003) report a $K$-band galaxy magnitude, 
which has been converted into a bulge magnitude by Graham \& Driver (2007, 
their Section 4.2.1) that we use here.  
While Gebhardt et al.\ (2003) provide a BH mass for this galaxy, it is
worth noting that 
their zero BH mass model is only ruled out at the 1.5-2$\sigma$ level.  
Lauer et al.\ (2005) provide a $V$-band estimate for 
the magnitude of the central nucleus 
(a previously unpublished result from Gebhardt et al.\ 1996). 
The absence of radio flux at 1.4 GHz, according to the NRAO VLA
Sky Survey\footnote{http://www.cv.nrao.edu/nvss/NVSSlist.shtml} 
(NVSS; Condon et al.\ 1998), suggests that the nuclear source 
is due to thermal emission from stars rather than non-thermal emission from an AGN. 
The stated $V-I$ colour in Lauer et al.\ (2005) is 1.3, corresponding to a range 
in the stellar $M/L$ ratio of $0.28 < \log [M/L_V] < 0.66$.
Given the similarity with the colour for the NC in NGC~1023, we adopt
$M/L_V = 2.5$, giving a NC mass of $1.1\times10^7 M_{\odot}$ when using our updated 
distance. 

\begin{figure}
\includegraphics[angle=270,scale=0.34]{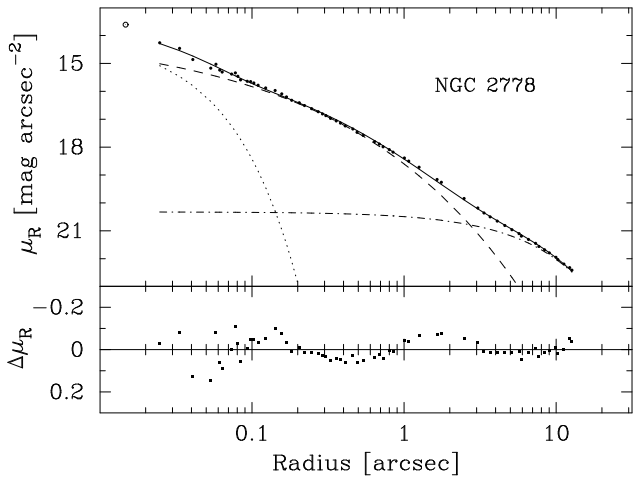}
\caption{
Deconvolved {\it F702W} ($R$-band) WFPC2, minor-axis surface brightness
profile for the galaxy NGC~2778 (Rest et al.\ 2001).  Dust filaments were
masked out prior to profile extraction, which is fitted 
here with an inner $n=1$ (NC) S\'ersic component, plus a S\'ersic bulge and 
an outer exponential disc. 
The bulge/disc nature of this galaxy is evident in 
the ground-based, minor-axis $r$-band light profile 
shown in Kent (1985, his Fig.2). 
The inner most component modelled here has an apparent $F702W$ 
magnitude of 20.0 mag.
}
\label{Fig_2778}
\end{figure}

Rest et al.\ (2001) did however report 
that there is no nuclear star cluster in NGC~2778 
--- although they noted that they had ``been fairly conservative in assigning
nucleation''.  We have therefore inspected their published 
F702W WFPC2 data, and model their (Dexter-extracted)\footnote{A hard disc
 crash caused the loss of the original data (Rest 2009, priv.\ comm.).}  
minor-axis light profile in our Figure~\ref{Fig_2778}.  Using the minor axis
light profile helps to avoid the discy, or bar-like, feature at $\sim$5
arcseconds ($\sim$0.5 kpc) which dominates both the major-axis light profile
and the ellipticity profile.  We find that the nuclear component, fitted with 
an $n=1$ S\'ersic model, has 
an apparent F702W magnitude of 20.0 mag.  Given the $V-I$ colour of 1.3, we have a
well constrained $M/L_R$ ratio of 1.50$^{+0.20}_{-0.15}$, 
giving a NC mass of $6.7\times10^6 M_{\odot}$, which we have used in this
study. This value is only 40 per cent smaller than the value derived from the 
$V$-band data.

{\bf NGC~3115:} 
This is another lenticular galaxy with a large-scale rotating disc (Rubin et
al.\ 1980).  
Kormendy et al.\ (1996) and Emsellem et al.\ (1999) have noted the
presence of a nuclear point source, while Lauer et al.\ (2005) provide a $V$-band
magnitude for this NC which equates to $-12.1$ mag for a galaxy 
distance of 9.7 Mpc (Tonry et al.\ 2001). 
%
The $V-I$ colour in Lauer et al.\ (2005) is 1.2, corresponding to
$0.18 < \log [M/L_V] < 0.56$ for all combinations of stellar metallicities and
ages.  We again adopt $M/L_V = 2.5$ for the NC.

{\bf NGC 3384:}
We have used the NC's $H$-band magnitude from Ravindranath et al.\ (2001). 
The nuclear $V-I$ colour of 1.3 from Lauer et al.\ (2005), 
implies a $H$-band $M/L$ ratio of $\sim0.60^{+0.25}_{-0.15}$ 
assuming [Fe/H] $<$ 0.56 (Figure~\ref{Fig_Lee}). 
From this we derive a NC mass of $2.2\times10^7 M_{\odot}$. 
As a check, we note that 
Lauer et al.\ (2005, their Table 8) provide a $V$-band NC
luminosity of $4\times10^6 L_{\odot}$.  Multiplying by an 
$M/L_V$ ratio of 2.5
gives a mass of $10^7 M_{\odot}$, similar within a factor of $\sim$2.

%

{\bf NGC 3621:}
This Sd galaxy contains an AGN and a NC (Satyapal et al. 2007; Barth et al.\
2008).  The bulge is evident as excess light over the inner 10 arcseconds of
the disc.

{\bf NGC 4026:} 
We have taken the black hole mass from G\"ultekin et al.\ (2009) and adjusted
its value according to the distance given by Tonry et al.\ (2001), who report
a distance modulus of 30.67 for this edge-on lenticular galaxy.
The $V$-band nucelar cluster magnitude is reported by Lauer et al.\ (2005,
their Table~8) to be $-11.6$ mag when using a distance of 15.6 Mpc; which we
adjust to $-11.3$ mag for our adopted distance of 13.6 Mpc. 
The innermost $V-I$ colour, shown in Figure~3 of Lauer et al.\ (2005) is 1.1,
corresponding to $M/L_V = 2\pm1$ (see Figure~\ref{Fig_Lee}).
The mean $V-I$ colour of the bulge is 1.3, indicative of a $2 < M/L_V < 4.5$. 
G\"ultekin et al.\ (2009) report an $M/L_V$ ratio of $4.5\pm0.3$
which we shall adopt here.  
The bulge magnitude is reported to be $M_V = -20.32$ mag (Lauer et al.\ 2005), 
which we adjust to $-20.0$ mag for our adopted distance.  
This has then been converted into a bulge magnitude by assuming
a typical lenticular galaxy bulge-to-total flux ratio of 1/4 (e.g., Graham \& Worley 2008, and
references therein), and achieved by adding $2.5\log(1/4)$ to give a value of
$-18.5$ mag. 


{\bf NGC 4564:} 
C\^ot\'e et al.\ (2006) identified this S0 galaxy as hosting both a NC
and a BH; although they were unable to measure the brightness of its NC. 
%
%
From the surface brightness, colour and ellipticity profile for NGC~4564 (VCC
1664) in Figure~61 from Ferrarese et al.\ (2006b), one can discern that this
S0 galaxy has a relatively blue ($g-z = 1.52$) nucleus and a large-scale
stellar disc which starts to dominate beyond $\sim$5 arcseconds.  However
no reliable structural decomposition is available. 
%

\begin{figure}
\includegraphics[angle=270,scale=0.34]{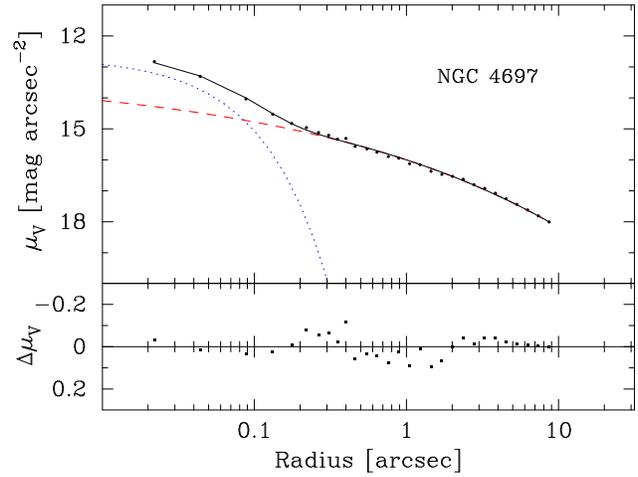}
\caption{
Deconvolved {\it F555W} ($V$-band) surface brightness profile for the galaxy
NGC~4697, taken from Byun et al.\ (1996) who remarked that any dust patches were 
masked out from the raw image before the profile extraction. 
The profile is fitted 
here with an inner $n=1$ (NC) S\'ersic component and an outer (galactic) 
$n=4$ S\'ersic component (Soria et al.\ 2006).  The inner component
has an apparent $F555W$ magnitude of 17.5 mag.
}
\label{Fig_4697}
\end{figure}

{\bf NGC 4697:} 
The nuclear dust disc in this galaxy was masked out prior to the extraction 
of the light profile (Byun et al.\ 1996). 
In Figure~\ref{Fig_4697} we have simultaneously 
fit the inner-component of NGC~4697's surface brightness profile, taken 
from Byun et al.\ (1996), with an $n=1$ S\'ersic model 
and the main galaxy with an $n=4$ S\'ersic model (Soria 
et al.\ 2006).  
This yields a magnitude for the inner component of 17.5 F555W-mag and 
a half-light radius of 4.4 parsecs. 
Using a distance modulus of 30.34 (Tonry et al.\ 2001) and $V-F555W = 0.0$
(Fukugita et al.\ 1995), one obtains an absolute magnitude of $-12.8$ $V$-mag.
%
%
Based on the properties of the preceding galaxies, we have adopted $M/L_V = 2.5$
for the NC.

{\bf NGC 7457:} 
Ravindranath et al.\ (2001) report an $H$-band magnitude 
for the nuclear cluster of NGC~7457 which is 0.8 mag brighter 
than the value in Balcells et al.\ (2007) because they treated 
the point source and the nuclear disc as a single entity. 
From a spectroscopic analysis, Sil'chenko et al.\ (2002) report a nuclear cluster 
age of 2-2.5 Gyr.  Together with the $V-I = 1.2$ colour from Lauer et al.\
(2005), this implies a metallicity of $\sim$0.5 
and therefore a $H$-band $M/L$ ratio of $\sim$0.4
(Figure~\ref{Fig_Lee}). 
%
Sil'chenko et al.\ (2002) also report a host bulge age $< 7$ Gyr, which is in 
accord with our adopted value for the bulges of disc galaxies. 

The BH mass for NGC~7457 was derived assuming that the excess central light,
above the inward extrapolation of the outer light profile, is AGN flux
(Gebhardt et al.\ 2003).  While Ho et al.\ (1995) see no
obvious nuclear emission from NGC~7457, Gebhardt et al.\ (2003) suggested that
it may be a weak BL Lac object.  Their BH mass determination inherently
assumed there is no additional nuclear mass components present.  However some
of the excess nuclear light emanates from what is a nuclear disc and a likely
star cluster (Balcells et al.\ 2007).  The BH mass is therefore
unfortunately in error at an unknown level.  Resolving this issue is beyond
the scope of this paper and we simply flag the data from this galaxy as
uncertain.
%
%

\subsection{Systems with only an upper limit to $M_{\rm BH}$}

{\bf NGC 598 (M33):}
For the Scd galaxy M33, we corrected the galaxy's apparent $B$-band magnitude
of 6.27 mag (RC3) for 0.18 mag of Galactic extinction (via NED), and using an
inclination of 54 degrees we corrected for 0.35 mag of internal extinction
(Driver et al.\ 2007).  This resulted in an absolute $B$-band magnitude of
$-18.78$ mag for this galaxy.  The average $B$-band bulge-to-total ($B/T$)
flux ratio for Scd galaxies is 0.027 (Graham \& Worley 2008) giving an
expected bulge magnitude of $-14.86$ mag.  From a bulge/disc decomposition of
M33, Bothun (1992) reports a comparable $B$-band bulge-to-disc ($B/D$) ratio
of 0.02 (see also Minniti et al.\ 1993; Regan \& Vogel 1994; and Mighell \&
Rich 1995).  
We do however note that while there is an obvious excess of flux
above the inward extrapolation of this galaxy's outer exponential light
distribution, M33 does not possess a traditional bulge (Wyse et al.\ 1997;
Brown 2009).  The central excess of stars might be better thought of as a
``pseudo-bulge'' rather than a classical bulge, i.e.\ a small elliptical
galaxy, 
%
and readers may ignore this system if they wish. 
The central bulge regions of M33 have been reported as 
8-10 Gyr old (Li et al.\ 2004), and 
Wyse et al.\ (1997) note $-2.2 <$ [Fe/H] $< -0.7$. This implies $-0.11 < \log
[M/L_B] < 0.17$, and we adopt the midpoint giving $M/L_B = 1.1$ for the
(pseudo-)bulge in M33. 

The NC mass of $2\times10^6 M_{\odot}$ was obtained from Kormendy
\& McLure (1993).  
NC masses should perhaps not be obtained from the virial theorem's
approximation, $M \approx \sigma^2R$, because they are not isolated systems,
but reside within the potential and pressure of their host spheroid.
%
%
As a rough check, we note that the nuclear $V-I$ colour is $0.85\pm0.05$ while
the galaxy $V-I$ colour is 1.0 (Gebhardt et al.\ 2001).  Schmidt et al.\
(1990) have reported intermediate to old stars, with [Fe/H] $> 0.1$, dominate
the NC, and so from a diagram of $M/L_B$ versus $V-I$ (akin to
Figure~\ref{Fig_Lee}) one has that $\log [M/L_B] = 0.27 \pm 0.22$ dex.  Using
the $B$-band NC magnitude of $-10.2$ mag from Kormendy \& McLure (1993), the
NC mass is $3\times10^6 M_{\odot}$.

{\bf NGC 205:} 

\begin{figure}
\includegraphics[angle=270,scale=0.34]{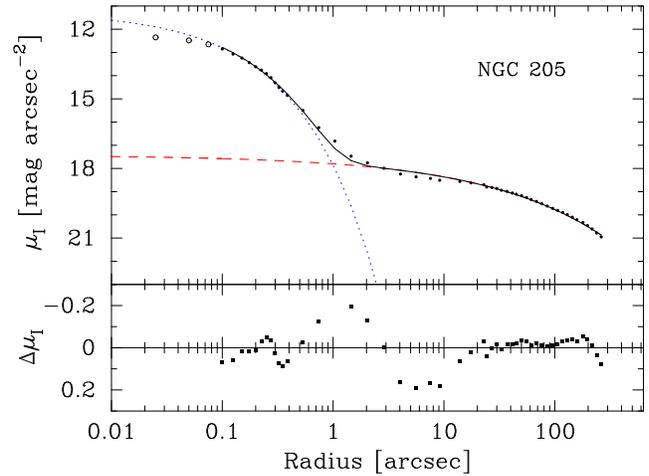}
\label{Fig_205} 
\caption{
Observed {\it HST/ACS F814W} (roughly Johnson $I$-band) surface brightness
profile for the galaxy NGC~205 from Valluri et al.\ (2005).  It is 
fitted here with an inner and outer S\'ersic component, both with free S\'ersic indices. 
Due to finite resolution, i.e.\ ``seeing'', the inner 3 points 
have been excluded from the fit.  The inner component has an apparent 
$F814W$ magnitude of 14.15 mag, an effective half-light radius equal to 0.3
arcseconds and a S\'ersic index equal to 1.6. 
}
\end{figure}

In addition to M33, NGC~205 also has only an upper limit to the mass of any
potential BH at its centre (Valluri et al.\ 2005).  It does however possess
a very obvious nuclear star cluster (e.g., Carter \& Sadler 1990).  From the
{\it F814W} surface brightness profile in Figure~\ref{Fig_205}, the inner
nuclear component has an absolute magnitude of $-10.42$ $I$-mag (using a
distance of 0.82 Mpc)\footnote{We elected not to use the nuclear cluster
  magnitudes from Butler \& Mart{\'{\i}}nez-Delgado (2005) because they were
  derived from Nuker model fits whose outer power-law results in excessive
  light at large radii being assigned to the NC.}.
%
%
There is some evidence for excess light from $\sim$0.5 to 3 arcseconds and
Butler \& Mart{\'{\i}}nez-Delgado (2005) have remarked that there may be a
nuclear disc present.
Using $M/L_I = 1.6(\times 0.95) = 1.52$ from Valluri et al.\ (2005), the
associated mass of the nuclear cluster is $0.8\times 10^6 M_{\odot}$, in fair 
agreement with our adopted (dynamical) value of $1.4\times 10^6 M_{\odot}$ 
from de Rijcke et al.\ (2006a).

%

For a 13 Gyr old spheroid population, $M/L_B \sim 2.5$, which is in good
agreement with our adopted value of 2.7 from De Rijcke et al.\ (2006a).

{\bf NGC 4041:} 
The absolute bulge magnitude has been derived from the
(extinction corrected) apparent galaxy magnitude $m_B = 11.8$ mag (RC3 and
NED) using the average, $B$-band, Sbc bulge-to-total flux ratio 0.069 (Graham \& Worley
2008), giving $m_{\rm bulge} = 14.7 B$-mag. 
This was transformed into a mass 
by first deriving a metallicity of 
[Fe/H] $\sim$ 0.25 dex, using $H-K \sim 0.25$ (2MASS)\footnote{Two micron all
  sky survey: Jarrett et al.\ (2000).} and $B-V = 0.67$ 
(RC3).   Assuming an age of 1-5 Gyr for the bulge, the associated 
$B$-band $M/L$ ratio ranges from 0.3 to 1.2.  We adopted the
midpoint of 0.6, and acknowledge a factor of two uncertainty in this value. 

The nuclear cluster mass was constrained using the photometry from Marconi et
al.\ (2003).  From their nuclear colours, $R-I = 0.6$ and $B-R = 1.5$, the 
logarithm of the $I$-band $M/L$ ratio is constrained to $0.0\pm0.1$,
and we adopt $M/L_I = 1.0$ for the NC. 
%

{\bf VCC 1254:} 
The upper limit to $M_{\rm BH}$ ($<9\times10^6 M_{\odot}$) 
is from Geha et al.\ (2002, their Section
3.2.3). 
It is however a rather high upper limit, as revealed by the location of this 
object in the $M_{\rm BH}$-$\sigma$ diagram given its velocity dispersion 
$\sigma=31$ kms s$^{-1}$. 
The nuclear cluster has a $V-I$ colour of $1.05\pm0.05$ 
while the main galaxy has colour gradient increasing from $\sim$1.2 at 1 arcsecond
to $\sim$1.40 at 15 arcseconds (Stiavelli et al.\ 2001). 
The galaxy's colour implies a metallicity [Fe/H] greater than $\sim -0.1$ dex. 
Assuming that the NC has a metallicity (between -0.1 and +0.56, see
Figure~\ref{Fig_Lee}), then its colour implies $-0.2 < \log [M/L_V] < +0.14$, 
and we have adopted $M/L_V = 0.93$ for the nucleus. 
From Figure~\ref{Fig_Lee}, we know that the galaxy must be relatively old, 
and adopting the outer $V-I$ colour, one has $M/L_V \sim 5-6$.  Geha
et al.\ (2002) reported $M/L_V$ = 6 and we have adopted this here for the
galaxy. 


\subsection{Globular and Star Clusters} 


For the first three globular clusters noted below, our estimates of their
stellar mass --- which we use to roughly check their adopted dynamical mass
--- are derived assuming that their stellar populations are predominantly old.

{\bf M15:}
NGC~7078 was initially heralded as the first detection of a GC with an 
intermediate mass black hole (IMBH) 
(Yanny et al.\ 1994; Gerssen et al.\ 2002, 2003).  
These claims have however since been questioned (de Paolis et al.\ 1996; 
Baumgardt et al.\ 2003a; van den Bosch et al.\ 2006; Bash et al.\ 2008).  The
absolute $V$-band magnitude from the updated Harris (1996) catalogue is
$-9.13$ mag (after dust correction).  Using the metallicity [Fe/H] = $-2.26$
dex from Harris (1996), one has $\log M/L_V \sim 0.3$ dex, giving a stellar
mass of $7.7\times10^5 M_{\odot}$.  This is in good agreement with our adopted
mass of $7.0\times10^5 M_{\odot}$ from Dull et al.\ (1997), after correcting
their distance of 7.2 kpc to 10.3 kpc (van den Bosch et al.\ 2006).

{\bf G1:}
Gebhardt et al.\ (2002, 2005) have claimed the existence of an IMBH in the
suspected globular cluster G1 (but see Ma et al.\ 2009) associated with the
Andromeda galaxy M31; 
with their zero mass BH model ruled out at the 97 per cent level. 
However a consensus has not been reached with plausible alternative scenarios 
yet to be properly ruled out (Baumgardt et al.\ 2003b; Pooley \& Rappaport 2006). 
Again assuming a predominantly old stellar population, and using the 
metallicity [Fe/H] = $-0.95$ dex from Harris (1996), 
one has $\log [M/L_V] \sim 0.3$ dex. 
Brightening the cluster's published absolute $V$-band magnitude 
(Meylan et al.\ 2001) by 0.1 mag for our new distance, and by 0.2 mag for 
Galactic extinction, its magnitude of $-11.24$ mag implies a stellar mass 
of $5.1\times10^6 M_{\odot}$, in reasonable agreement with our adopted 
value of $8.0\times10^6 M_{\odot}$ (Baumgardt et al.\ 2003b).

{\bf $\omega$ Cen:} 
As noted by Noyola et al.\ (2008), this object may 
have an IMBH, although these authors conclude that ``detailed
numerical simulations are required to confidently rule out other
possibilities''.  While we have adopted the dynamical mass for the cluster of
$4.7\times10^6 M_{\odot}$ from Meylan et al.\ (1995), we note that our
derivation of the stellar mass is $\sim$2.5 times smaller.  The absolute
$V$-band magnitude from the updated Harris (1996) catalogue is $-10.13$ mag 
(after dust correction).  
Using the metallicity from Harris [Fe/H] = $-1.62$ dex 
gives $\log M/L_V \sim 0.3$ dex and 
thus a stellar mass of $1.9\times10^6 M_{\odot}$.  

Dynamical-to-stellar mass ratios of 2 to 3 are not uncommon for bright
globular clusters (e.g., Dabringhausen et al.\ 2008; Forbes et al.\ 2008).  It
is thought that this may be a reflection that the adopted initial stellar mass
function is inappropriate, which therefore affects the estimated stellar $M/L$
ratio.  The multiple stellar populations in $\omega$ Cen (e.g., Stanford et
al.\ 2007, and references therein), and its possible origin as the nucleus of
a stripped dwarf galaxy (Bekki \& Norris 2006; Georgiev et al.\ 2009) also
complicate matters.

{\bf RZ2109, 47 Tuc, NGC 6388, NGC 6752, and RBS 1032:} 
The globular cluster RZ2109 associated with the Virgo elliptical galaxy NGC 4472
is not included as claims for its IMBH appear to have been premature (Zepf et
al.\ 2008).   
The globular cluster 47 Tuc is also excluded (de Rijcke et al.\ 2006b;
McLaughlin et al.\ 2006a) as the 1$\sigma$ uncertainties on the IMBH mass 
are consistent with a value of zero. 
Similarly, due to uncertainties as to their existence, potential 
intermediate mass BHs in 
NGC 6388 (Lanzoni et al.\ 2007; Nucita et al.\ 2008), 
NGC 6752 (Ferraro et al.\ 2003) and 
RBS 1032 (Ghosh et al.\ 2006) are also not included at this time. 
%
Lists of additional (lesser known) globular clusters which have been probed by
others for signs of an IMBH can be seen in Maccarone \& Servillat (2008, their Table~1).

{\bf MGG 11:}
The potential IMBH within the dense, young star cluster MGG-11
(Portegies Zwart et al.\ 2004; Patruno et al.\ 2006) --- 
which is only $\sim$200 pc from the centre of NGC~3034 (M82) --- 
is the brightest X-ray source in M82 .  
Dynamical friction (e.g., Bellazzini et al.\ 2008) is expected to result in the eventual 
centralisation of this star cluster within M82, albeit after some probable growth
to the IMBH and perhaps some evaporation of the star cluster (e.g.. 
Ebisuzaki et al.\ 2001). 
The available BH mass estimate has however been acquired from a rather 
indirect method, and we refer readers to the cautionary note 
in Berghea et al.\ (2008) when considering the validity of this object.

The brightest ultraluminous X-ray source in the southern ring of the Cartwheel
galaxy, known as N.10, has not yet been mated with any specific 
star cluster and hence, even if it is an intermediate mass BH, 
cannot be included (Wolter et al.\ 2006).  Similarly, no such association
is known for MCG03-34-63 X-1 (Miniutti et al.\ 2006) which is also excluded. 
Other ultraluminous X-ray sources, such as X-1 in the dwarf irregular galaxy
Holmberg II (Miller et al.\ 2005), are also excluded due to their uncertain
nature (Berghea et al.\ 2008).

\subsection{Active Galactic Nuclei}

An increasing number of galaxies with both NCs and active galactic nuclei
(AGN) have been found.  Perhaps the most well known is NGC~4395 (Filippenko \&
Ho 2003) because useful estimates of its BH mass exist.
A second example is NGC~4303 (Colina et al.\ 2002), however its BH mass is 
poorly constrained (Pastorini et al.\ 2007) and thus not included here. 
From an analysis of 176 galaxies previously identified to have a NC, Seth et
al.\ increase the count by identifying ten (their section 3.3: 
NGC~1042, NGC~3259, NGC~4411B, 
NGC~4750, NGC~5377, NGC~5879, NGC~6000, NGC~6384, NGC~6951, NGC~7418) 
plus two (their section 4: NGC~4321, NGC~5921). 
Furthermore, an additional $\sim$30 galaxies have been identified as having
some indication of a possible BH.
However, without useful estimates of their BH masses, we can only include a
fraction of such systems.

{\bf NGC 4395:}
An inspection of Filippenko \& Ho's (2003)
Figure~3 suggests that the unaccounted for excess flux seen from 1 to $\sim$2
arcseconds, peaking at $\sim$0.6 mag arcsec$^{-2}$ above their fitted
model, may be the (expectedly small) bulge of this Sd spiral galaxy.  Starting
with the total apparent $B$-band galaxy magnitude of 10.64 mag (RC3),
we have applied a Galactic dust correction $A_B=0.07$ mag (Schlegel et
al.\ 1998) and, assuming an inclination of 34 degrees, a (disc) 
inclination-attenuation correction of 0.25 mag (Driver et al.\ 2007).
We have then used the typical $B$-band $B/T$ flux ratio for Sd
galaxies ($=0.027$, Graham \& Worley 2008) to obtain a dust-corrected
bulge magnitude of 14.24 $B$-mag.  At a distance of 4.3 Mpc, the absolute
$B$-band bulge magnitude is $-13.93$ mag.
To derive the bulge mass, we first derived a metallicity [Fe/H] = +0.2, using
the 2MASS colour H-K = 0.25 and the RC3 colour B-V = 0.45.  Assuming an age of
1-5 Gyr for the bulge, the associated B-band mass-to-light ratio ranges from
0.3 to 1.2.  We have adopted the midpoint of 0.6, and acknowledge a factor of
two uncertainty in this value.

The NC mass has been obtained assuming an $I$-band mass-to-light
ratio of 1.0, following NGC~4041, which has then been applied to the $I$-band
magnitude (Filippenko \& Ho 2003).
%
%
Using an array of methods, the BH mass for NGC~4395 is estimated
to be $4 < \log M_{\rm BH} < 5$ (Filippenko \& Ho 2003).
We have adopted the value $\log M_{\rm BH} = 4.5$ dex.  We do however
note that subsequent reverberation mapping estimates (Peterson et al.\ 2005) 
predicts a greater mass of $3.6\pm1.1 \times 10^5 M_{\odot}$.

{\bf NGC 1042:}
The Scd galaxy NGC~1042 has a NC (B\"oker et al.\ 2003), a bulge (Knapen et
al.\ 2003) with a dust corrected magnitude of $-19.13$ $K$-mag (Graham \& Worley
2008), and a BH (Seth et al.\ 2008, see also Shields et al.\ 2008).  
From the $K$-band $M_{\rm BH}$-$L_{\rm
  bulge}$ relation in Graham (2007), the expected BH mass is $3\times 10^6
M_{\odot}$, equal to the NC's stellar mass (Walcher et al.\ 2005). 
However, a direct measurement of the BH mass is not known and so we do not
include this system.

\subsection{Additional galaxies}

We have included 13 ``core'' galaxies to help illustrate/define the high-mass end
of the various diagrams/relations.  In particular, as already noted, we have 
used NGC~1399: the brightest cluster galaxy from Fornax.  
From a cursory inspection of the many {\it HST}-resolved surface brightness profiles
available in the literature, one will discover several core galaxies 
that appear to be nucleated.  However, one must distinguish 
between excess central flux from star clusters and the 
non-thermal emission from AGN. 
For example, Ravindranath et al.\ (2001) reveal a point source in NGC~4374,
but this is a LINER Sy2 AGN.
It is however noted here that, like NGC~1399, NGC~4649 
displays a central $\sigma$-drop 
(Pinkey et al.\ 2003) which might be associated with a nuclear
stellar cluster, but in any event the BH will dominate the central mass budget. 

{\bf NGC 4552:}
While Faber et al.\ (1997) report neither severe nor moderate nucleation in
the core galaxy 
NGC~4552, a small amount of excess flux within 0$\arcsec$.1 may be apparent in
the light profiles shown by Byun et al.\ (1996) and Carollo et al.\ (1997).
Renzini et al.\ (1995) has however reported that the point source in this
LINER Seyfert 2 galaxy is variable at UV wavelengths, and is therefore likely to be
due to an AGN.  We have however elected not to include this S0 galaxy 
(Caon et al.\ 1993) because an accurate spheroid mass is not availble. 

{\bf IC 1459:} 
Although Lauer et al.\ (2005) provide a $V$-magnitude for the nucleus of
IC~1459, the blue point source (Forbes et al.\ 1995; Tomita et al.\ 2000;
Verdoes Kleijn et al.\ 2002) is a LINER associated with an AGN that has been
detected in X-rays (Fabbiano et al.\ 2003).  We have excluded this
object, due to the the order of magnitude 
uncertainty on its BH mass (Cappellari et al.\ 2002).

{\bf NGC 2748:} The apparent nuclear star cluster is due to dust (Hughes et
al.\ 2005; Seigar et al.\ 2002). 
%

%

{\bf M31}
NGC~224, better known as Andromeda, has not been included because the 
nucleus is not a pressure supported star cluster in which the stars 
have random motions but is a rotationally supported nuclear disc. 



\begin{table*}
\caption{Black hole, host spheroid and nuclear cluster masses}
\label{TabS1}
\begin{tabular}{@{}lcccccccc@{}}
\hline
Object   &   Dist.   & $M_{\rm bh}$                          & Mag$_{\rm spheroid}$ &  $M/L$    & Mass$_{\rm spheroid}$   &  Mag$_{\rm NC}$     & $M/L$    & Mass$_{\rm NC}$       \\
         &  [Mpc]    & [$M_{\odot}$]                         &     [mag]            &           &  [$M_{\odot}$]          &    [mag]            &          & [$M_{\odot}$]         \\
1        &   2       &   3                                   &       4              &     5     &     6                   &      7              &     8    &   9                   \\
\hline
\multicolumn{9}{c}{Systems with $M_{\rm bh}$ but no detectable NC} \\
NGC 3379 &  10.6     & $1.4^{+2.7}_{-1.0}\times10^8$ [10]    & $-24.2\, K$-mag [35] &  1.0      &  $1.0\times10^{11}$     &  ...                & ...      & ...                  \\
NGC 3608 &  22.9     & $1.9^{+1.0}_{-0.6}\times10^8$ [11]    & $-24.1\, K$-mag [35] &  1.0      &  $9.4\times10^{10}$     &  ...                & ...      & ...                  \\
NGC 4261 &  31.6     & $5.2^{+1.0}_{-1.1}\times10^8$ [12]    & $-25.6\, K$-mag [35] &  1.0      &  $3.7\times10^{11}$     &  ...                & ...      & ...                  \\
NGC 4291 &  26.2     & $3.1^{+0.8}_{-2.3}\times10^8$ [11]    & $-23.9\, K$-mag [35] &  1.0      &  $7.8\times10^{10}$     &  ...                & ...      & ...                  \\
NGC 4374 &  18.4     & $4.6^{+3.5}_{-1.8}\times10^8$ [13]    & $-25.7\, K$-mag [35] &  1.0      &  $4.1\times10^{11}$     &  ...                & ...      & ...                  \\
NGC 4473 &  15.7     & $1.1^{+0.4}_{-0.8}\times10^8$ [11]    & $-23.8\, K$-mag [35] &  1.0      &  $7.1\times10^{10}$     &  ...                & ...      & ...                  \\
NGC 4486 &  16.1     & $3.4^{+1.0}_{-1.0}\times10^9$ [14]    & $-25.6\, K$-mag [35] &  1.0      &  $3.7\times10^{11}$     &  ...                & ...      & ...                  \\
NGC 4649 &  16.8     & $2.0^{+0.4}_{-0.6}\times10^9$ [11]    & $-25.8\, K$-mag [35] &  1.0      &  $4.5\times10^{11}$     &  ...                & ...      & ...                  \\
NGC 5077 &  41.2 [1] & $7.4^{+4.7}_{-3.0}\times10^8$ [15]    & $-21.1\, B$-mag [36] &  2.5      &  $1.1\times10^{11}$     &  ...                & ...      & ...                  \\	 
NGC 5813 &  32.2     & $7.0^{+1.1}_{-1.1}\times10^8$ [16]    & $-21.4\, B$-mag [36] &  2.5      &  $1.4\times10^{11}$     &  ...                & ...      & ...                  \\	 
NGC 6251 &  105 [1]  & $5.9^{+2.0}_{-2.0}\times10^8$ [17]    & $-26.6\, K$-mag [35] &  1.0      &  $9.4\times10^{11}$     &  ...                & ...      & ...                  \\
NGC 7052 &  66.4 [1] & $3.7^{+2.6}_{-1.5}\times10^8$ [18]    & $-21.6\, B$-mag [37] &  2.5      &  $1.7\times10^{11}$     &  ...                & ...      & ...                  \\  
\multicolumn{9}{c}{Systems with $M_{\rm bh}$ and a NC} \\ 
Milky Way &  0.008   & $3.7^{+0.2}_{-0.2}\times10^6$ [19]   &   ...                 &  ...      & $1.2\times10^{10}$ [46] &      ...              &  ...   & $3.0\times10^7$ [49] \\ 
M32       &  0.8     & $2.5^{+0.5}_{-0.5}\times10^6$ [20]   &  $-17.0\, I$-mag [38] &  0.9      &  $2.6\times10^{8}$      &  $-14.6\, I$-mag [38] &  0.75  & $2.0\times10^7$  \\        
NGC 1023 & 11.4      & $4.4^{+0.5}_{-0.5}\times10^7$ [21]   &  $-23.5\, K$-mag [35] &  0.6      &  $3.2\times10^{10}$     &  $-10.8\, V$-mag [21,41] & 2.5 & $4.4\times10^6$     \\  
NGC 1399 & 20.0      & $4.8^{+0.7}_{-0.7}\times10^8$ [22]   &  $-21.5\, B$-mag [39] &  2.5      &  $1.5\times10^{11}$     &  $-11.2\, V$-mag [47] &  2.5   & $6.4\times10^6$      \\ 
NGC 2778 & 22.9      & $1.4^{+0.8}_{-0.9}\times10^7$ [11]   &  $-21.3\, K$-mag [40] &  0.6      &  $4.3\times10^9$        &  $-11.8\, R$-mag [38] &  1.5   & $6.7\times10^6$      \\ 
NGC 3115 &  9.7      & $9.1^{+9.9}_{-2.8}\times10^8$ [23]   &  $-24.4\, K$-mag [35] &  0.6      &  $7.4\times10^{10}$     &  $-12.1\, V$-mag [41] &  2.5   & $1.5\times10^7$      \\ 
NGC 3384 & 11.6      & $1.6^{+0.1}_{-0.2}\times10^7$ [11]   &  $-22.6\, K$-mag [35] &  0.6      &  $1.4\times10^{10}$     &  $-15.6\, H$-mag [48] &  0.6   & $2.2\times10^7$      \\ 
NGC 4026 & 13.6      & $1.8^{+0.6}_{-0.4}\times10^8$ [24]   &  $-18.5\, V$-mag [41] &  4.5      &  $9.6\times10^9$        &  $-11.3\, V$-mag [41] &  2.0   & $5.6\times10^6$      \\  
NGC 4395 & 4.3  [2]  & $3.2^{+6.8}_{-2.2}\times10^4$ [25]   &  $-13.9\, B$-mag [37] &  0.6      &  $3.4\times10^7$        &  $-11.4\, I$-mag [25] & 1.0    & $1.4\times10^6$      \\  
NGC 4564 & 15.0      & $5.6^{+0.3}_{-0.8}\times10^7$ [11]   &  $-21.9\, K$-mag [40] &  0.6      &  $7.4\times10^{9}$      &        ?              &   ?    &   ?                  \\  
NGC 4697 & 11.7      & $1.7^{+0.2}_{-0.1}\times10^8$ [11]   &  $-24.6\, K$-mag [35] &  1.0      &  $1.5\times10^{11}$     &  $-12.8\, V$-mag [38] &  2.5   & $2.8\times10^7$      \\  
NGC 7457 & 13.2      & $3.5^{+1.1}_{-1.4}\times10^6$ [11]   &  $-19.8\, K$-mag [42] &  0.6      &  $1.1\times10^{9}$      &  $-15.1\, H$-mag [42] & 0.4    & $9.3\times10^6$      \\  
\multicolumn{9}{c}{Systems with only an upper limit on $M_{\rm bh}$ but with a NC} \\
M33      & 0.8 [3]   &  $<3\times10^3$ [26]                 &  $-14.9\, B$-mag [37,38] & 1.1    &  $1.5\times10^8$        &     ...               &  ...   & $2\times10^6$ [50]   \\  
NGC 205  & 0.82 [4]  &  $<2.4\times10^4$ [27]               &  $-15.8\, B$-mag [43] & 2.7 [45]  &  $8.7\times10^8$        &     ...               & ...    & $1.4\times10^6$ [45] \\  
NGC 3621 & 6.6       &  $<3.6\times10^6$ [28]               &  $-17.6\, K$-mag [28] &  0.6      &  $1.4\times10^{8}$      &   ...                 &  ...   & $1.0\times10^7$ [28]  \\ 
NGC 4041 & 23.3 [1]  &  $<2.4\times10^7$ [29]               &  $-17.1\, B$-mag [37,38] & 0.6    &  $6.4\times10^8$        &  $-14.7\, I$-mag [29] & 1.0    & $2.9\times10^7$      \\  
VCC 1254 & 17.0 [5]  &  $<9\times10^6$ [30]                 &  $-17.0\, V$-mag [44] & 6.0 [30]  &  $3.2\times10^9$        &  $-12.9\, V$-mag [44] & 0.93   & $1.1\times10^7$      \\  
\multicolumn{9}{c}{Systems with less secure $M_{\rm bh}$ but with a NC}\\
G1       &  0.8  [6]  & $1.8^{+0.5}_{-0.5}\times10^4$ [31]  &      ...              &  ...      &          ...            &       ...             &   ...  & $8.0\times10^6$ [51] \\ 
M15      & 0.01 [7]   & $0.5^{+2.5}_{-0.5}\times10^3$ [32]  &      ...              &  ...      &          ...            &       ...             &   ...  & $7.0\times10^5$ [52] \\ 
MGG-11   &  3.6 [8]   & $1.0^{+4.0}_{-0.8}\times10^3$ [33]  &      ...              &  ...      &          ...            &       ...             &   ...  & $3.5\times10^5$ [8]  \\ 
$\omega$ Cen & 0.0048 [9] & $4.0^{+0.75}_{-1.0}\times10^4$ [34] &  ...              &  ...      &          ...            &       ...             &   ...  & $4.7\times10^6$ [53] \\ 
\hline
\end{tabular}
\noindent
Column~1: In addition to galaxies, three globular clusters (G1, M15 and $\omega$ Cen) and one star cluster (MGG-11) are listed. 
Column~2: Distances have been taken from Tonry et al.\ (2001) unless otherwise noted.
Column~3: BH masses have been adjusted to the distances shown in column~2. 
Column~4: Absolute magnitude of the host spheroid using the distances in column~2. 
Column~5: Stellar mass-to-light ratio (associated with the filter specified in column~4) used to obtain the stellar mass of the spheroid (Column~6).
Column~7: Absolute magnitude of the nuclear star cluster (or globular cluster) using the distances in column~2. 
Column~8: Stellar mass-to-light ratio (associated with the filter specified in column~7) used to obtain the stellar mass of the cluster (Column~9).

{\bf References:} 
1 = NED (Virgo + GA + Shapley)-corrected Hubble Flow distance); 
2 = Thim et al.\ (2004);
3 = Argon et al.\ (2004);
4 = McConnachie et al.\ (2005);   
5 = Jerjen et al.\ (2004);   
6 = the Tonry et al.\ (2001) distance to NGC~224 (M31) is used; 
7 = Harris (1996); 
8 = McCrady et al.\ (2003); 
9 = van de Ven et al.\ (2006); 
10 = Shapiro et al.\ (2006); 
11 = Gebhardt et al.\ (2003);
12 = Ferrarese et al.\ (1996); 
13 = Maciejewski \& Binney (2001);
14 = Macchetto et al.\ (1997);
15 = De Francesco et al.\ (2008); 
16 = Preliminary values determined by Hu (2008) from Conf.\ Proc.\ figures of Cappellari et al.\ (2008);
17 = Ferrarese \& Ford (1999); 
18 = van der Marel \& van den Bosch (1998); 
19 = Ghez et al.\ (2005);
20 = Verolme et al.\ (2002); 
21 = Bower et al.\ (2001);
22 = Houghton et al.\ (2006), Gebhardt et al.\ (2007);
23 = Emsellem et al.\ (1999);   
24 = G\"ultekin et al.\ (2009); 
25 = Filippenko \& Ho (2003);   
26 = Merritt et al.\ (2001),  Gebhardt et al.\ (2001); 
27 = Jones et al.\ (1996), Valluri et al.\ (2005); 
28 = Barth et al.\ (2008); 
29 = Marconi et al.\ (2003):
30 = Geha et al.\ (2002);
31 = Gebhardt et al.\ (2005) but see Baumgardt et al.\ (2003b); 
32 = Gerssen et al.\ (2003), van den Bosch et al.\ (2006) but see Baumgardt et al.\ (2003a) and Dull et al.\ (2003); 
33 = Patruno et al.\ (2006); 
34 = Noyola et al.\ (2008);  
35 = Marconi \& Hunt (2003);
36 = Rest et al.\ (2001); 
37 = de Vaucouleurs et al.\ (1991, RC3); 
38 = This paper;
39 = D'Onofrio et al.\ (1994); 
40 = Graham \& Driver's (2007, their Section~4.2.1) bulge magnitude; 
41 = Lauer et al.\ (2005, their table~8); 
42 = Balcells et al.\ (2007); 
43 = Mateo (1998); 
44 = Geha et al.\ (2003); 
45 = De Rijcke et al.\ (2006a);
46 = Cardone \& Sereno (2005), Dwek et al.\ (1995); 
47 = Lyubenova et al.\ (2008);
48 = Ravindranath et al.\ (2001); 
49 = Launhardt et al.\ (2002); Sch\"odel et al.\ (2007); 
50 = Kormendy \& McLure (1993);  
51 = Baumgardt et al.\ (2003b); Ma et al.\ (2009);
52 = Dull et al.\ (1997); 
53 = Meylan et al.\ (1995).
%
\end{table*}

\label{lastpage}
\end{document}